\documentclass[12pt]{article}
\usepackage{a4wide} 

\RequirePackage[latin1]{inputenc}    

\usepackage[english]{babel}
\usepackage{amsmath}
\usepackage{amsfonts}
\usepackage{amssymb}
\usepackage{xspace}
\usepackage{latexsym}
\usepackage{url}
\usepackage{xspace}
\usepackage{fancyvrb}

\usepackage{graphics,color}             
\RequirePackage[latin1]{inputenc}  

\newenvironment{restate-proposition}[2][{}]{\noindent\textbf{Proposition~{#2}}\;\textbf{#1}\ 
}{\vskip 1em}

\newenvironment{restate-theorem}[2][{}]{\noindent\textbf{Theorem~{#2}}\;\textbf{#1}\ 
}{\vskip 1em}

\newenvironment{restate-corollary}[2][{}]{\noindent\textbf{Corollary~{#2}}\;\textbf{#1}\ 
}{\vskip 1em}

\newcommand{\Proofitemb}[1]{\medskip \noindent {\bf #1\;}}
\newcommand{\Proofitemfb}[1]{\noindent {\bf #1\;}}
\newcommand{\Proofitem}[1]{\medskip \noindent $#1\;$}
\newcommand{\Proofitemf}[1]{\noindent $#1\;$}
\newcommand{\Defitem}[1]{\smallskip \noindent $#1\;$}
\newcommand{\Defitemt}[1]{\smallskip \noindent {\em #1\;}}
\newcommand{\Defitemf}[1]{\noindent $#1\;$}

\makeatother

\makeatletter
\def\@ysproof[#1]{\@sproof{ #1}}
\def\@sproof#1{\begin{trivlist}\item[]{\textit{Sketch of the proof#1.}}}

\makeatletter
\def\@yproof[#1]{\@proof{ #1}}
\def\@proof#1{\begin{trivlist}\item[]{\textit{Proof#1.}}}

\newcommand{\hbra}{\noindent\hbox to \textwidth{\leaders\hrule height1.8mm depth-1.5mm\hfill}}
\newcommand{\hket}{\noindent\hbox to \textwidth{\leaders\hrule height0.3mm\hfill}}
\newcommand{\ratio}{.3}

\newtheorem{theorem}{Theorem}

\newtheorem{definition}[theorem]{Definition}

\newtheorem{corollary}[theorem]{Corollary}
\newtheorem{proposition}[theorem]{Proposition}

\newtheorem{remark}[theorem]{Remark}

\newcommand{\Proof}{\noindent {\sc Proof}. }
\newcommand{\Proofhint}{\noindent {\sc Proof hint}. }
\newcommand{\qed}{\hfill${\Box}$}

\newcommand{\Figbar}{{\center \rule{\hsize}{0.3mm}}}    

\newcommand{\cl}[1]{{\cal #1}}          

\newcommand{\ul}[1]{\underline{#1}}     
\newcommand{\ol}[1]{\overline{#1}}      

\newcommand{\arrow}{\rightarrow}        
\newcommand{\Alt}{ \mid\!\!\mid  }
\newcommand{\isum}{\oplus}

\newcommand{\infer}[2]{\begin{array}{c} #1 \\ \hline #2 \end{array}}

\newcommand{\Arrow}{\Rightarrow}        

\newcommand{\union}{\cup}               
\newcommand{\inter}{\cap}               
\newcommand{\minus}{\backslash}         
\newcommand{\comp}{\circ}               
\newcommand{\set}[1]{\{#1\}}            

\newcommand{\dcl}{\downarrow}           
\newcommand{\ucl}{\uparrow}             
\newcommand{\rl}[1]{\;{\cal #1}\;}             
\newcommand{\rel}[1]{{\cal #1}}         

\newcommand{\fn}[1]{{\it fn}(#1)}                       

\newcommand{\mand}{\mbox{ and }}
\newcommand{\w}[1]{{\it #1}}    

\newcommand{\xst}[2]{\exists\, #1\;\: #2}

\newcommand{\s}[1]{{\sf #1}}    
\newcommand{\vc}[1]{{\bf #1}}

\newcommand{\act}[1]{\stackrel{#1}{\rightarrow}} 
\newcommand{\wact}[1]{\stackrel{#1}{\Rightarrow}} 

\newcommand{\iact}[1]{\stackrel{#1}{\longmapsto}} 

\newcommand{\susp}{\downarrow}
\newcommand{\lsusp}{\Downarrow_L}
\newcommand{\wsusp}{\Downarrow}
\newcommand{\commits}{\searrow}

\newcommand{\spi}{S\pi}

\newcommand{\bbis}{\approx_{B}}
\newcommand{\cbis}{\approx_{C}}
\newcommand{\lbis}{\approx_{L}}

\newcommand{\bbissusp}{\approx_{B}^{\susp}}
\newcommand{\cbissusp}{\approx_{C}^{\susp}}
\newcommand{\lbissusp}{\approx_{L}^{\susp}}

\newcommand{\bbiswsusp}{\approx_{B}^{\wsusp}}
\newcommand{\cbiswsusp}{\approx_{C}^{\wsusp}}
\newcommand{\lbiswsusp}{\approx_{L}^{\wsusp}}

\newcommand{\sbis}{\equiv_L}
\newcommand{\emit}[2]{\ol{#1}#2} 
\newcommand{\present}[4]{#1(#2).#3,#4}
\newcommand{\match}[4]{[#1=#2]#3,#4}       
\newcommand{\matchv}[4]{[#1 \unrhd #2]#3,#4}  
\newcommand{\new}[2]{\nu #1 \ #2}
\newcommand{\outact}[3]{\new{{\bf #1}}{\emit{#2}{#3}}}
\newcommand{\real}{\makebox[5mm]{\,$\|\!-$}}

\begin{document}

\title{A synchronous $\pi$-calculus }

\author{Roberto M. Amadio\thanks{Partially supported by ANR-06-SETI-010-02.}\\
        Universit\'e Paris 7\thanks{Laboratoire {\em Preuves,
        Programmes et Syst\`emes}, UMR-CNRS 7126.}}

\maketitle

\begin{abstract}
The SL synchronous programming model 
is a relaxation of the {\sc Esterel} synchronous model where
the reaction to the absence of a signal within an instant can 
only happen at the next instant. 
In previous work, we have revisited the SL
synchronous programming model. In particular, we have discussed an
alternative design of the model,
introduced a CPS translation to a tail
recursive form, and proposed a notion of bisimulation
equivalence.
In the present work, we extend the tail recursive model with
first-order data types obtaining a non-deterministic
synchronous model whose complexity is comparable to 
the one of the $\pi$-calculus.  We show that
our approach to bisimulation equivalence can cope with
this extension and in particular that labelled bisimulation
can be characterised as a contextual bisimulation.
\end{abstract}

\newpage

\section{Introduction}
Concurrent and/or distributed systems are usually classified according
to two main parameters (see, {\em e.g.}, \cite{LL90}): the {\em
relative speed} of the processes (or threads, or components,
or nodes) and their {\em interaction mechanism}. With respect to the
first parameter one refers to synchronous, asynchronous, partially
synchronous,$\ldots$ systems. In particular, in {\em synchronous
systems}, there is a notion of {\em instant} (or phase, or pulse, or
round) and at each instant each process performs some actions and
synchronizes with all other processes. One may say that
all processes proceed at the same speed and it is in this specific sense
that we will refer to {\em synchrony} in this work.

With respect to the second parameter, one considers shared memory,
message passing, signals, broadcast,$\ldots$ Concerning the message
passing interaction mechanism, one distinguishes various situations
according to whether the communication channel includes a bounded or
unbounded and an ordered or unordered buffer.  In particular the
situation where the buffer has $0$ capacity corresponds to a {\em
rendez-vous communication} mechanism which is also called {\em
synchronous communication} in that it forces a synchronisation.

The notion of synchrony (in the sense adopted in this work)
is a valuable logical concept that simplifies
the design and analysis of systems. One may verify 
this claim by consulting standard textbooks in concurrent/distributed 
algorithms such as \cite{Lynch96,Tel94} and comparing the algorithms for
basic problems such as leader election, minimum spanning tree, 
consensus,$\ldots$ in the synchronous and asynchronous case.
In \cite{Lynch96,Tel94}, the formalisation of the so called {\em synchronous 
network model} is quite simple. One assumes a fixed network topology
and describes the behaviour of each process essentially as an infinite
state Moore machine \cite{HU79}: at each instant, each process, depending on
its current state, emits a message on each outgoing edge, then
it receives a messages from each incoming edge, and computes its
state for the next instant.

In this paper, we are looking at the synchronous model from the point
of view of {\em process calculi}. This means in particular, that we 
are looking for a notion of equivalence of synchronous systems with
good compositionality properties.
The works on SCCS \cite{Milner83} and Meije \cite{AB84} are an early
attempt at providing a process calculus representation of the
synchronous model. SCCS and Meije are built over the same action
structure: essentially, the free abelian group generated by a set of
{\em particulate} actions. The models differ in the choice of the
combinators: SCCS starts with a {\em synchronous} parallel composition
and then adds operators to {\em desynchronise} processes while Meije
starts with an {\em asynchronous} parallel composition and then adds
operators that allow to synchronise processes.  As a matter of fact,
the SCCS and Meije operators are inter-definable so that the calculi
can be regarded as two presentations of the same model.  

SCCS/Meije is a simple model with nice mathematical properties but
it has failed so far to turn into a model for a realistic synchronous
programming language. For this reason, we will not take the SCCS/Meije
model as a starting point, but the {\em synchronous language} SL
introduced in \cite{BD95}.  
Threads in the SL model interact through {\em signals} as opposed to
channels. A cooperative scheduling (as opposed to pre-emptive, see
\cite{Ous96}) is sometimes considered, though this is not quite a
compulsory choice and it is not followed here.  
This style of synchronous and possibly cooperative
programming has been advocated as a more effective approach to the
development of applications such as event-driven controllers, data flow
architectures, graphical user interfaces, simulations, web services,
multi-player games (we refer to \cite{ABBC05} for a discussion of the
applications and implementation techniques).

The SL model can be regarded as a relaxation of the {\sc Esterel}
model \cite{BG92} where the reaction to the {\em absence} of a signal within an
instant can only happen at the next instant.  This design choice
avoids some paradoxical situations and simplifies the
implementation of the model.
Unlike the SCCS/Meije model, the SL model has gradually evolved into a
general purpose programming language for concurrent applications and
has been embedded in various programming environments such as
\textsc{C}, \textsc{Java}, \textsc{Scheme}, and \textsc{Caml} (see
\cite{boussinot:rc91,mimosarp,SchemeFT,MandelPouzetPPDP05}).  For
instance, the Reactive ML language \cite{MandelPouzetPPDP05} includes
a large fragment of the \textsc{Caml} language plus primitives to
generate signals and synchronise on them.  We should also mention that
related ideas have been developed by Saraswat et al. \cite{SJG96} in
the area of constraint programming.

The Meije and the {\sc Esterel}/SL models were developed in
Sophia-Antipolis in the same research team, but, as of today, there
seems to be no strong positive or negative result on the possibility
of representing one of the models into the other. Still there are a
number of features that plead in favour of the {\sc Esterel}/SL
model. First, the shift from channel based to signal based
communication allows to preserve (to some extent) the {\em
determinacy} of the computation while allowing for multi-point
interaction. Second, pure signals, {\em i.e.}, signals carrying no
values, as opposed to pure channels, allow for a representation of data
in binary rather than unary notation.  Third, there is a natural
generalisation of the calculus to include general data types.
Fourth, the length of an instant is {\em programmable} 
rather than being given {\em in extenso} as a finite word of 
so called {\em particulate actions}. Fifth, efficient
implementations of the model have been developed.

In the early 80's, the development of the SCCS/Meije model relied on
the {\em same} mathematical framework (labelled transition system and
bisimulation) that was used for the development of the CCS model. 
However, the following years have witnessed the
development of two quite distinct research directions concerned with
asynchronous and synchronous programming, respectively.  Nowadays, the
$\pi$-calculus \cite{MPW92} and its relatives can be regarded as
typical abstract models of asynchronous concurrent programming while
various languages such as {\sc Lustre} \cite{CPHP87}, {\sc Esterel}
\cite{BG92}, and SL \cite{BD95} carry 
the flag of synchronous programming. 

We remark that while the $\pi$-calculus has inherited many of the
techniques developed for CCS, the semantic theory of the SL model
remains largely underdeveloped.  In recent work
\cite{Amadio05}, we have revisited the SL synchronous programming
model. In particular, we have discussed an alternative design of the
model,
introduced a CPS translation to a tail recursive form, and proposed a
novel notion of bisimulation equivalence with good compositionality
properties.  The original SL language as well as the revised one
assume that signals are {\em pure} in the sense that they carry no
value.  Then computations are naturally deterministic and bisimulation
equivalence collapses with trace equivalence.  However, practical
programming languages that have been developed on top of the model
include data types beyond {\em pure signals} and this extension makes
the computation {\em non-deterministic} unless significant
restrictions are imposed.  For instance, in the Reactive ML language
we have already quoted, signals carry values and the emission of two
distinct values on the same signal may produce a non-deterministic
behaviour.

In the present work, we introduce a minimal extension of the tail recursive
model where signals may carry first-order values including signal
names.  The linguistic complexity of the resulting language is
comparable to the one of the $\pi$-calculus and we tentatively call it
the $\spi$-calculus (pronounced $s-pi$).\footnote{S for {\em
synchronous} as in SCCS \cite{Milner89} and SL \cite{BD95}.  
Not to be confused with the so called `synchronous' $\pi$-calculus 
which would be more correctly
described as the $\pi$-calculus with {\em rendez-vous} communication
nor with the SPI-calculus where the S suggests a pervasive `spy'
controlling and corrupting all communications.}
Our contribution is to show that the notion of bisimulation
equivalence introduced in \cite{Amadio05} is sufficiently robust to be
lifted from the deterministic language with pure signals to the
non-deterministic language with data types and signal name generation.
The main role in this story is played by a new notion of labelled
bisimulation.
We show that this notion has good congruence properties and that it
can be characterised via a suitable notion of contextual bisimulation
in the sense of \cite{HY95}.  The proof of the characterisation
theorem turns out to be considerably more complex than in the pure
case having to cope with phenomena such as non-determinism and name
extrusion.  

While this approach to the semantics of concurrency has already been
explored in the framework of asynchronous languages including, {\em
e.g.}, the $\pi$-calculus \cite{HY95,ACS98,FG98}, Prasad's calculus of
broadcasting systems \cite{P95,HR98}, and the ambient calculus
\cite{MZN05}, this seems to be the first concrete application of the
approach to a {\em synchronous} language.  We expect that the
resulting semantic theory for the SL model will have a positive
fall-out on the development of various static analyses techniques to
guarantee properties such as {\em determinacy}
\cite{MandelPouzetPPDP05}, {\em reactivity} \cite{AD04}, and {\em
non-interference} \cite{BCM04}.  

In the following, we assume familiarity with the technical development
of the theory of bisimulation for the $\pi$-calculus and some
acquaintance with the synchronous languages of the {\sc Esterel}
family.

\section{The $\spi$-calculus}
Programs $P,Q,\ldots$ in the $\spi$-calculus
are defined as follows:
\[
\begin{array}{ll}
P &::= 0 \Alt A(\vc{e}) \Alt \emit{s}{e} \Alt
\present{s}{x}{P}{K} 
\Alt \match{s_1}{s_2}{P_1}{P_2}
\Alt \matchv{u}{p}{P_1}{P_2}
\Alt \new{s}{P}
\Alt P_1\mid P_2 \\
K &::=A(\vc{r})
\end{array}
\]
We use the notation $\vc{m}$ for a vector $m_1,\ldots,m_n$, $n\geq 0$.
The informal behaviour of programs follows.
$0$ is the terminated thread. $A(\vc{e})$ is a (tail) recursive call
with a vector $\vc{e}$ of expressions as argument. The identifier $A$
is defined by a unique equation $A(\vc{x})=P$ with the usual
condition that the variables free in $P$ are contained in $\set{\vc{x}}$.
$\emit{s}{e}$ evaluates the expression $e$ and emits its value on the
signal $s$. A value emitted on a signal {\em persists} within the
instant and it is {\em reset} at the end of each instant. 
 $\present{s}{x}{P}{K}$ is the {\em present} statement
which is the fundamental operator of the SL model. If the values
$v_1,\ldots,v_n$ have been emitted on the signal $s$ in the current instant 
then $\present{s}{x}{P}{K}$ evolves
non-deterministically into $[v_i/x]P$ for some $v_i$ ($[\_/\_]$ is our
notation for substitution).  On the other
hand, if no value is emitted then the continuation $K$ is evaluated at
the end of the instant.  $\match{s_1}{s_2}{P_1}{P_2}$ is the usual
matching function of the $\pi$-calculus 
that runs $P_1$ if $s_1=s_2$ and $P_2$, otherwise.
Here both $s_1$ and $s_2$ are free.
$\matchv{u}{p}{P_1}{P_2}$, matches $u$ against the pattern $p$.
We assume $u$ is either a variable $x$ or a value $v$ and $p$ has
the shape $\s{c}(\vc{p})$, where $\s{c}$ is a constructor and 
$\vc{p}$ a vector of patterns. 
At run time, $u$ is always a {\em value} and we 
run $\sigma P_1$ if $\sigma$
is the result of matching $u$ against $p$, and $P_2$ otherwise.
Note that as usual the variables occurring in the pattern $p$ are bound.
$\new{s}{P}$ creates a new signal name $s$ and runs $P$.
$(P_1\mid P_2)$ runs in parallel $P_1$ and $P_2$.  The continuation $K$
is simply a recursive call whose arguments are either expressions
or values associated with signals at the end of the instant in
a sense that we explain below.\footnote{The reader may have noticed 
that we prefer the term {\em program} to  the term {\em process}. 
By this choice, we want to stress that the parallel threads that
compose a program are tightly coupled and are executed and observed as a whole.}

The definition of program relies on the following syntactic categories:
\[
\begin{array}{lll}
\w{Sig} &::= s \Alt t \Alt \cdots  &\mbox{(signal names)} \\
\w{Var} &::= \w{Sig} \Alt x \Alt y \Alt z \Alt \cdots   &\mbox{(variables)} \\
\w{Cnst} &::= \s{*} \Alt \s{nil} \Alt \s{cons} \Alt \s{c} \Alt \s{d} \Alt\cdots &\mbox{(constructors)} \\
\w{Val} &::= \w{Sig} \Alt \w{Cnst}(\w{Val},\ldots,\w{Val})
&\mbox{(values $v,v',\ldots$)}\\
\w{Pat} &::= \w{Var} \Alt \w{Cnst}(\w{Pat},\ldots,\w{Pat})
&\mbox{(patterns $p,p',\ldots$)} \\
\w{Exp} &::=\w{Pat} 
&\mbox{(expressions $e,e',\ldots$)} \\
\w{Rexp} &::= !\w{Sig} \Alt \w{Var} \Alt \w{Cnst}(\w{Rexp},\ldots,\w{Rexp})
&\mbox{(exp. with dereferenciation $r,r',\ldots$)} 
\end{array}
\]
As in the $\pi$-calculus, signal names stand both for
signal constants as generated by the $\nu$ operator and signal
variables as in the formal parameter of the present operator.
Variables $\w{Var}$ include signal names as well as variables of other
types.  Constructors $\w{Cnst}$ include $\s{*}$, $\s{nil}$, and
$\s{cons}$. We will also write $[v_{1};\ldots;v_{n}]$ for the
list of values $\s{cons}(v_{1},\ldots,\s{cons}(v_{n},\s{nil})\ldots)$,
$n\geq 0$. Values $\w{Val}$ are terms built out of constructors and signal names.
Patterns $\w{Pat}$ are terms built out of constructors and variables
(including signal names).  
For the sake of simplicity, expressions $\w{Exp}$ 
here happen to be the same as patterns
but we could easily add first-order functional symbols defined by
recursive equations.
Finally, $\w{Rexp}$ is composed of either expressions or
the dereferenced value of a signal at the
end of the instant. Intuitively, the latter
corresponds to the set of values emitted on 
the signal during the instant.
If $P, p$ are a program and a pattern then  we denote
with $\w{fn}(P), \w{fn}(p)$ the set of free signal names
occurring in them, respectively. We also use $\w{FV}(P), \w{FV}(p)$
to denote the set of free variables (including signal names).

\subsection{Typing}
Types include  the basic type $1$ inhabited by the constant $*$ and, 
assuming $t$ is a type,  the type $\w{sig}(t)$ of signals carrying
values of type $t$, and the type $\w{list}(t)$ of lists of values of
type $t$ with constructors \s{nil} and \s{cons}.
$1$ and $\w{list}(t)$ are examples of {\em inductive types}. More
inductive types (booleans, numbers, trees,$\ldots$) 
can be added along with more constructors.  
We assume that variables (including signals), constructor
symbols, and thread identifiers come with their (first-order) types. 
For instance, a constructor $\s{c}$ may have a type
$(t_1,t_2)\arrow t$ meaning that it waits two arguments of
type $t_1$ and $t_2$ respectively and returns a value of type $t$.
It is then straightforward to define when a program is well-typed 
and verify that this property is preserved by the following 
reduction semantics. 
We just notice that if a signal name $s$ has type $\w{sig}(t)$ then its
dereferenced value $!s$ should have type $\w{list}(t)$. In the
following, we will tacitly assume that we are handling well typed
programs, expressions, substitutions,$\ldots$

\subsection{Matching}
As already mentioned, the $\spi$-calculus includes two distinct matching
constructions: one operating over signal names works as in the
$\pi$-calculus and the other operating over values of inductive type
actually computes a matching substitution $\w{match}(v,p)$ which
is defined as follows:\footnote{Without loss of expressive power, 
one could assume that in the second
matching instruction the pattern $p$ 
contains exactly one constructor symbol and that all the variables
occurring in it are distinct.}
\[
\w{match}(v,p) =\left\{
\begin{array}{ll}
\sigma   &\mbox{if } \w{dom}(\sigma)=\w{FV}(p),\quad \sigma(p)=v \\
\ucl     &\mbox{otherwise}
\end{array}\right.
\]
To appreciate the difference, assume $s\neq s'$ and consider
$P= \match{s}{s'}{P_1}{P_2}$ and 
$P'=\matchv{[s]}{[s']}{P_1}{P_2}$.
In the first case, $P$ reduces to $P_2$
while in the second case, $P'$ reduces to $[s/s']P_1$.
Indeed, in the first case $s'$ is
a constant while in the second case 
it is a bound variable.

\subsection{Informal reduction semantics}
Assume $v_1\neq v_2$ are two distinct values and consider the
following program in $\spi$:
\[
\begin{array}{l}
P=\nu \ s_1,s_2 \ 
(\quad \emit{s_{1}}{v_{1}} \quad \mid \quad 
 \emit{s_{1}}{v_{2}} \quad \mid  \quad 
 s_1(x). \ (s_1(y). \  (s_2(z). \ A(x,y) \  \ul{,B(!s_1)}) \quad
 \ul{,0}) \quad \ul{,0} \quad )
\end{array}
\]
If we forget about the underlined parts and we regard $s_1,s_2$ as
{\em channel names} then $P$ could also be viewed as a $\pi$-calculus
process. In this case, $P$ would reduce to 
\[
P_1 =  \new{s_1,s_2}{(s_2(z).A(\sigma(x),\sigma(y))}
\]
where $\sigma$ is a substitution such that
$\sigma(x),\sigma(y)\in \set{v_1,v_2}$ and $\sigma(x)\neq \sigma(y)$.
In $\spi$, {\em signals persist within the instant} and 
$P$ reduces to 
\[
P_2 = \new{s_1,s_2}{(\emit{s_{1}}{v_{1}} \mid 
 \emit{s_{1}}{v_{2}} \mid (s_2(z).A(\sigma(x),\sigma(y)),B(!s_1)))}
\]
where  $\sigma(x),\sigma(y)\in \set{v_1,v_2}$.

One can easily formalise this behaviour by assuming a standard
structural equivalence, by introducing the
usual rules for matching and for unfolding recursive definitions 
(cf. rules  $=_{1}^{\w{sig}}$, $=_{2}^{\w{sig}}$, $=_{1}^{\w{ind}}$,
$=_{2}^{\w{ind}}$, and $\w{rec}$ in the following Table \ref{lts}), and by 
adding the rule:
\[
\emit{s}{v} \mid s(x).P,K \arrow \emit{s}{v} \mid [v/x]P
\]
What happens next? In the $\pi$-calculus, $P_1$ is 
{\em deadlocked} and no further computation is possible.
In the $\spi$-calculus, 
the fact that no further computation is possible in $P_2$ is
detected and marks the {\em end of the current instant}. Then
an additional computation represented by the relation $\mapsto$ 
moves $P_2$ to the following instant:
\[
P_2 \mapsto P'_2 =  \new{s_1,s_2}{B(\ell)}
\]
where $\ell \in  \set{[v_1;v_2],[v_2;v_1]}$.
Thus at the end of the instant, a dereferenced signal such as $!s_{1}$
becomes a list of (distinct) values emitted on $s_1$ during the
instant and then all signals are reset. 

We will further comment on the relationships between the
$\pi$-calculus 
and the $\spi$-calculus in section \ref{comparison} 
once the formal definitions are in
place. In the following section \ref{transition}, 
Table \ref{lts} will formalise 
the reduction relation (in the special
case where the transition is labelled with the action $\tau$) while
Table \ref{endofinst} will describe the evaluation relation at the end of
the instant.

\subsection{Transitions}\label{transition}
The behaviour of a program 
is specified by (i) a {\em labelled transition
system} $\act{\alpha}$ describing the possible interactions of the program
{\em during an instant} and (ii) a {\em transition system} $\mapsto$ determining
how a program evolves at the {\em end of each instant}. 

As usual, the behaviour is defined only for programs 
whose only free variables are signals.
The labelled transition system is similar to the one of 
the polyadic $\pi$-calculus modulo a different
treatment of emission which we explain below.
We define actions $\alpha$ as follows:
\[
\alpha::= \tau \Alt sv \Alt \outact{t}{s}{v}
\]
where in the emission action the signal names $\vc{t}$ are distinct,
occur in $v$, and differ from $s$.  The functions
$n$ (names), $\w{fn}$ (free names), and $\w{bn}$ (bound names) are
defined on actions as usual:
$\w{fn}(\tau)=\emptyset$, $\w{fn}(sv)=\set{s}\union \w{fn}(v)$,
$\w{fn}(\outact{t}{s}{v})= (\set{s}\union \w{fn}(v))\minus
\set{\vc{t}}$;
$\w{bn}(\tau)=\w{bn}(sv)=\emptyset$,
$\w{bn}(\outact{t}{s}{v})=\set{\vc{t}}$; 
$\w{n}(\alpha)=\w{fn}(\alpha)\union \w{bn}(\alpha)$.
The related labelled transition system 
is defined in table \ref{lts} where rules apply only to
programs whose only free variables are signal names and with
standard conventions on the renaming of bound names.
As usual, the symmetric rule for $(\w{par})$ and $(\w{synch})$
are omitted.
\begin{table}
\[
\begin{array}{cc}

(\w{out})\quad
\infer{~}{\emit{s}{v} \act{\emit{s}{v}} \emit{s}{v}}

&(\w{in})\quad
\infer{~}{\present{s}{x}{P}{K} \act{sv} [v/x]P\mid \emit{s}{v}}
\\ \\

(\w{par})\quad
\infer{P_1\act{\alpha} P'_1 \quad \w{bn}(\alpha)\inter\w{fn}(P_2)=\emptyset}
{P_1\mid P_2\act{\alpha} P'_1\mid P_2}

&(\w{synch})\quad
\infer{P_1 \act{\outact{t}{s}{v}} P'_1\quad P_2 \act{sv}P'_2\quad
\set{\vc{t}}\inter \w{fn}(P_2)=\emptyset}
{P_1\mid P_2 \act{\tau} \new{\vc{t}}{(P'_1\mid P'_2)}} \\ \\

(\nu)\quad \infer{P\act{\alpha} P' \quad t\notin n(\alpha)}
{\new{t}{P}\act{\alpha} \new{t}{P'}}

&(\nu_{\w{ex}})\quad
\infer{P\act{\outact{t}{s}{v}} P'\quad t'\neq s\quad t'\in n(v)\minus \set{\vc{t}}}
{\new{t'}{P}\act{(\nu t',\vc{t})\emit{s}{v}} P'} \\ \\

(=_{1}^{\w{sig}})\quad \infer{~}{\match{s}{s}{P_1}{P_2}\act{\tau} P_1}

&(=_{2}^{\w{sig}})\quad \infer{s_1\neq s_2}{\match{s_1}{s_2}{P_1}{P_2}\act{\tau} P_2} \\ \\

(=_{1}^{\w{ind}})\quad \infer{\w{match}(v,p)=\sigma}{\matchv{v}{p}{P_1}{P_2}\act{\tau} \sigma P_1}

&(=_{2}^{\w{ind}})\quad \infer{\w{match}(v,p)=\ucl}{\matchv{v}{p}{P_1}{P_2} \act{\tau} P_2} \\ \\

(\w{rec})\quad \infer{A(\vc{x})=P}{A(\vc{v})\act{\tau}[\vc{v}/\vc{x}]P} 

\end{array}
\]
\caption{Labelled transition system during an instant}\label{lts}
\end{table}
The rules are those of the polyadic $\pi$-calculus
but for the following points. 
(1) In the rule $(\w{out})$, the emission is {\em persistent}. 
(2) In the rule $(\w{in})$, the continuation carries the memory
that the environment has emitted $\emit{s}{v}$. For example, this guarantees, 
that in the program $s(x).(s(y).P,0),0$, if the environment provides
a value $\emit{s}{v}$ for the first input then that value persists
and is available for the second input too.
(3) The rules $(=_{1}^{\w{ind}})$ and $(=_{2}^{\w{ind}})$ handle the
pattern matching. 
We write $P\act{\alpha} \cdot$ for $\xst{P'}{P\act{\alpha} P'}$.
We will also write  $P\wact{\tau} P'$ for $P (\act{\tau})^* P'$ and
$P\wact{\alpha} P'$ with $\alpha\neq \tau$ for
$P (\wact{\tau})$ $(\act{\alpha})(\wact{\tau}) P'$.

A program is {\em suspended}, {\em i.e.}, 
it reaches the {\em end of an instant}, when the labelled transition
system  cannot produce further (internal) $\tau$ transitions.

\begin{definition}
We write $P\susp$ if $\neg (P\act{\tau} \cdot)$ and
say that the program $P$ is suspended. 
\end{definition}

When the program $P$ is suspended, an additional
computation is carried on to move to the next instant.
This computation  is described by the transition system $\mapsto$.
First of all, we have to compute the set of values emitted
on every signal. To this end, we introduce some notation.

Let $E$ vary over functions from signal names
to finite sets of values. Denote with $\emptyset$
the function that associates the empty set with every
signal name,  with $[M/s]$ the function that associates 
the set $M$ with the signal name $s$ and the empty set
with all the other signal names, and with $\union$ the 
union of functions defined pointwise.

We represent a set of values as a list of the
values contained in the set. More precisely,
we write $v \real M$ and say that $v$ {\em represents} $M$ 
if $M=\set{v_1,\ldots,v_n}$ and
$v=[v_{\pi(1)};\ldots; v_{\pi(n)}]$ for
some permutation $\pi$ over $\set{1,\ldots,n}$.
Suppose $V$ is a function from signal names to lists of values.
We write $V\real E$ if $V(s)\real E(s)$ for every signal name $s$.
We also write $\w{dom}(V)$ for 
$\set{s \mid V(s)\neq []}$.
If $K$ is a continuation, {\em i.e.}, a recursive call $A(\vc{r})$,
then $V(K)$ is obtained from $K$ by replacing
each occurrence $!s$ of a dereferenced signal with the associated
value $V(s)$. We denote with $V[\ell/s]$ the function that behaves as
$V$ except on $s$ where $V[\ell/s](s)=\ell$.

To define the transition $\mapsto$ at the end of the instant, we rely
on an auxiliary judgement $P \iact{E,V} P'$.
Intuitively, this judgement states that:
(1) $P$ is suspended,
(2) $P$ emits exactly the values specified by $E$, and
(3) the behaviour of $P$ in the following instant is $P'$ and depends
on  $V$.
\begin{table}
\[
\begin{array}{c}

(0)\ \infer{~}
{0 \iact{\emptyset,V} 0}

\qquad
(\w{out})\ \infer{v \mbox{ occurs in }V(s)}
{\emit{s}{v} \iact{[\set{v}/s],V} 0}

\qquad
(\w{in})\ \infer{s\notin \w{dom}(V)}
{s(x).P,K \iact{\emptyset,V} V(K)} \\ \\

(\w{par})\ 
\infer{P_i \iact{E_{i},V} P'_i \quad i=1,2 }
{(P_1\mid P_2) \iact{E_1\union E_2,V} (P'_1\mid P'_2)}

\qquad

(\nu)\ 
\infer{P \iact{E,V'} P'\quad V'(s)\real E(s)\quad V[[]/s]=V'[[]/s]}
{\new{s}{P} \iact{E[\emptyset/s],V} \new{s}{P'}} \\\\

\infer{P \iact{E,V} P'\quad V\real E}
{P\mapsto P'}

\end{array}
\]
\caption{Transition system at the end of the instant}\label{endofinst}
\end{table}
The transition system presented in table  \ref{endofinst} 
formalizes this intuition. For instance, one can show that:
\[
\new{s_{1}}{(s_{1}(x).0,A(!s_{2}) \mid \emit{s_{2}}{v_{3}})} 
\mid (\emit{s_{2}}{v_{2}} \mid \emit{s_{1}}{v_{1}}) 
\iact{E,V}
\new{s_{1}}{(A(V(s_{2})) \mid 0)} 
\mid (0 \mid 0)
\]
where $E=[\set{v_{1}}/s_{1},\set{v_{2},v_{3}}/s_2]$ and, {\em e.g.},
$V=[[v_{1}]/s_{1},[v_{3};v_{2}]/s_{2}]$.

\subsection{Derived operators}\label{derived-op}
We introduce some derived operators and some abbreviations.
The calculi with {\em pure signals} considered in 
\cite{BD95,ABBC05,Amadio05} can
be recovered by assuming that all signals have type
$\w{Sig}(1)$. In this case, we will simply write $\ol{s}$ for
$\emit{s}{\s{*}}$ and $s.P,K$ for 
$\present{s}{x}{P}{K}$ where $x\notin\w{FV}(P)$.
We denote with $\Omega$ a {\em looping process} defined, {\em e.g.}, by
$\Omega=A()$ where $A() = A()$.
We abbreviate $\present{s}{x}{P}{0}$ with $s(x).P$.
We  can derive an {\em internal choice operator} by defining,
\[
P_1 \isum P_2
= 
\new{s}{(s(x)\matchv{x}{\s{0}}{P_1}{P_2} \mid \emit{s}{\s{0}} 
\mid \emit{s}{\s{1}})}
\]
where, {\em e.g.}, we set $\s{0}=[]$ and $\s{1}=[\s{*}]$.
The $\s{pause}$ operation suspends the execution till the end
of the instant. It is defined by:
\[
\s{pause}.K  = \new{s}{s.0,K}
\]
where: $s\notin \w{fn}(K)$. We can also 
simulate an operator $\s{await} \ s(x).P$ that waits for a
value on  a signal $s$ for arbitrarily many instants by defining:
\[
\s{await}\ s(x).P = 
\present{s}{x}{P}{A(\vc{x})}
\]
where $\set{\vc{x}}=\set{s}\union (\w{FV}(P)\minus \set{x})$ and 
$A(\vc{x})=\present{s}{x}{P}{A(\vc{x})}$.

It is also interesting to program a {\em generalised matching
operator} $[x=\new{\vc{s}}{v}]_X P$ that given a value $x$, checks
whether $x$ has the shape $\new{\vc{s}}{v}$ where the freshness of the
signal names $\vc{s}$ is relative to a finite set $X$ of signal names,
{\em i.e.}, no name in $\vc{s}$ belongs to $X$. If
this is the case, we run $P$ and otherwise we do nothing.
Assuming, $\set{\vc{s}}\subseteq \fn{v}$,
$\fn{v}\minus \set{\vc{s}}\subseteq X$, 
$\set{\vc{s}}\inter X=\emptyset$, and $X=\emptyset$ whenever
$\set{\vc{s}}=\emptyset$, there are three cases to consider:
\begin{enumerate}

\item $v=s$ is a signal name and $\vc{s}$ is empty. 
Then $[x=s]_X P$ is coded as $\match{x}{s}{P}{0}$.

\item $v=s$ is a signal name and $\vc{s}=s$. Then
$[x=\new{s}{s}]_X P$ is coded as $[x\notin X]P$ where if
$X=\set{s_1,\ldots,s_n}$ then 
$[x\notin X]P$ is coded as $[x=s_1]0,(\cdots,[x=s_n]0,P\cdots)$.

\item $v=\s{c}(p_1,\ldots,p_n)$. 
Let $\set{\vc{s'}}=\w{fn}(v)\minus  \set{\vc{s}}$
be the set of signal names which are free in $\new{\vc{s}}{v}$. 
We associate with the vector of signal names $\vc{s'}$ a vector of 
fresh signal names $\vc{s''}$. Let $v''=[\vc{s''}/\vc{s'}]v$.
Then $[x=\new{\vc{s}}{v}]_X P$ is coded as:
\[
[x\unrhd v''][\vc{s''}=\vc{s'}][\set{\vc{s}}\inter X=\emptyset ][\vc{s}
  \mbox{ distinct}]P
\]
where: (1) $[s''_1,\ldots,s''_m = s'_1,\ldots,s'_m]Q$ 
is an abbreviation
for $[s''_1=s'_1]\ldots ([s''_m=s'_m]Q,0)\ldots,0$,
(2) $[\set{\vc{s}}\inter X=\emptyset]$ is expressed by requiring
that every signal name in $\set{\vc{s}}$ does not belong to
$X$, and
(3) $[\vc{s} \mbox{ distinct}]$ is expressed by requiring that
the signal names in $\vc{s}$ are pairwise different.
For example, to express
\[
   [x=\new{s_1,s_2}{\s{c}(s_1,\s{c}(s'_3,s_2,s_1),s'_3)}]_{\set{s'_3,s'_4}}P
\]
we write
$[x \unrhd \s{c}(s_1,\s{c}(s''_3,s_2,s_1),s''_3)][s''_3=s'_3]
   [s_1\notin \set{s'_3,s'_4}][s_2\notin \set{s'_3,s'_4}][s_1\neq s_2]P$.
Note that the introduction of the auxiliary signal names $\vc{s''}$ is
required because in the pattern considered the signal names
are interpreted as variables and not as constants. Also, note
that the names $s_1$, $s_2$, and $s''_{3}$ are bound in $P$.
\end{enumerate}

\subsection{Comparison with the $\pi$-calculus}\label{comparison}
In order to make a comparison easier, 
the syntax of the $\spi$-calculus is similar to the one of
the $\pi$-calculus. However there are some important semantic
differences to keep in mind.

\Defitemt{Deadlock vs. End of instant.}
What happens when all threads are either terminated or
waiting for an event that cannot occur? In the $\pi$-calculus,
the computation stops. In the $\spi$-calculus (and more generally,
in the SL model), this  situation is detected and marks 
the end of the current instant.
Then suspended threads are  reinitialised, signals are 
reset, and the computation  moves to the following instant.

\Defitemt{Channels vs. Signals.}
In the $\pi$-calculus, a message is consumed by its recipient.
In the $\spi$-calculus, a value emitted along the 
signal persists within an  instant and it is reset at the end of it.
We note that in the semantics the only relevant information is 
whether a given value was emitted or not, {\em e.g.},
we do not distinguish the situation where the same value is emitted
once or twice within an instant.

\Defitemt{Data types.}
The (polyadic) $\pi$-calculus has {\em tuples} as basic data type, 
while the $\spi$-calculus has {\em lists}. 
The reason for including lists rather than tuples in the
{\em basic} calculus is that at the end of the instant we transform a
set of values into a suitable data structure (in our case a list) that
represents the set and that can be processed as a whole 
in the following instant. Note in particular, that the list associated
with a signal is empty if and only if no value was emitted on the
signal during the instant. This allows to detect the {\em absence} of 
a signal at the end of the instant.

\Defitemt{Determinism vs. Non-determinism.}
In the $\spi$-calculus there are two sources of non-determinism.
(1) Several values emitted on the same signal compete to be received 
during the instant, {\em e.g.},
$ \ol{s}\s{0} \mid \ol{s}\s{1} \mid s(x).P$ may evolve
into either $\emit{s}{\s{0}} \mid \emit{s}{\s{1}} \mid [\s{0}/x]P$ or
$\emit{s}{\s{0}} \mid \emit{s}{\s{1}} \mid [\s{1}/x]P$.
(2) At the end of the instant, values emitted on a signal are collected
in an order that cannot be predicted, {\em e.g.},
$\new{s',s''}(\ol{s} s' \mid \ol{s} s'' \mid \s{pause}.A(!s,s',s''))$ may 
evolve into either
$A([s';s''],s',s'')$ or
$A([s'';s'],s',s'')$.
Accordingly, one may consider 
two restrictions to make the computation {\em deterministic}.
(i) If a signal can be read {\em during} an instant then at most
one value can be emitted on that signal during an 
instant.\footnote{For instance, the calculus with pure signals 
satisfies this condition.}
(ii) If a signal can only be read {\em at the end} of the instant then
the processing of the associated list of values 
is {\em independent} of its order.\footnote{In the languages of the {\sc Esterel} family,
sometimes one makes the hypothesis that the values collected at the
end of the instant are combined by means of 
an associative and commutative function.
While this works in certain cases, it seems hard to conceive 
such a function when manipulating objects such as pointers. 
It seems that a general notion of deterministic program should be 
built upon a suitable notion of program equivalence such as the one we
develop here.}

\subsection{Comparison with CBS and the timed $\pi$-calculus}
In the {\em calculus of broadcasting systems} (CBS, \cite{P95}), 
threads interact through 
a {\em unique} broadcast channel. The execution
mechanism {\em guarantees} that at each step one process 
sends a message while all the other processes either receive the
message or ignore it. There is a similarity between the
emission of a value on a signal and the broadcast of a value in the sense
that in both cases the value can be received an arbitrary number of
times. On the other hand, it appears that the CBS model does
not offer a {\em direct} representation of the notion of instant.

Berger's timed $\pi$-calculus \cite{Berger04} includes a primitive $\s{timer}^t \
x(y).P,Q$ which means: wait for a message on $x$ for at most $t$ time
units and if it does not come then do $Q$. While there is a syntactic
similarity with the present statement of the SL model, we 
remark that the notion of {\em time unit} is very different from the
notion of {\em instant} in the SL model.  In the SL model, an instant
lasts exactly the time needed for every process to accomplish the
tasks it has scheduled for the current instant. In the timed model, a
time unit lasts exactly one reduction step. As a matter of fact, 
the notion of `reduction step' is based  on a rather arbitrary 
definition and it fails to be a {\em robust} programming concept.

\section{Labelled bisimulation and its characterisation}
We introduce a new notion of labelled bisimulation, a
related notion of contextual bisimulation and state
our main result: the two bisimulations coincide.

\begin{definition}\label{wl-susp-def}
We write: 
\[
\begin{array}{lll}
P \wsusp &\mbox{if } \xst{P'}{( \ P\wact{\tau} P' \mand P'\susp \ )}
&\mbox{(weak suspension)} \\ 
P\lsusp 
&\mbox{if } \xst{\alpha_{1},P_1\ldots,\alpha_{n},P_{n}}{( \ P\act{\alpha_{1}} P_1 \cdots \act{\alpha_{n}}P_n},
\quad n\geq 0, \mand P_n\susp \ ) 
&\mbox{(L-suspension)}
\end{array}
\]
\end{definition}

Obviously, $P\susp$ implies $P\wsusp$ which in turn implies 
$P\lsusp$ and we will see that these implications cannot be reversed.
The L-suspension predicate (L for labelled) plays
an important role in the definition of 
labelled bisimulation which is the central concept of this paper.

\begin{definition}[labelled bisimulation]\label{lab-bis-def}
A symmetric relation $\rel{R}$ on programs is a labelled bisimulation
if whenever 
$P\rl{R} Q$ the following holds:

\Defitem{(L1)}
If $P\act{\tau} P'$ then
$\xst{Q'}{(Q\wact{\tau} Q' \mand P'\rl{R}Q')}$.

\Defitem{(L2)}
If $P\act{\outact{t}{s}{v}}P'$, $P\lsusp$, 
$\set{\vc{t}}\inter \w{fn}(Q)=\emptyset$ then
$\xst{Q'}{(Q\wact{\outact{t}{s}{v}} Q' \mand P'
  \rl{R}Q')}$.

\Defitem{(L3)}
If $P\act{sv} P'$ then 
$\xst{Q'}{(\ (\ Q\wact{sv} Q'\mand P'\rl{R} Q') \mbox { or }
(\ Q\wact{\tau} Q'\mand  P'\rl{R} (Q'\mid \emit{s}{v}) \ ) \ )}$.

\Defitem{(L4)}
If $S= \emit{s_{1}}{v_{1}} \mid \cdots \mid \emit{s_{n}}{v_{n}}$,
$n\geq 0$, 
$P'= (P\mid S) \susp$, and $P'\mapsto P''$
then\\ \noindent
$\xst{Q',Q''}{(\ (Q \mid S) \wact{\tau} Q',\ 
Q'\susp,\  P'\rl{R} Q', \ Q'\mapsto Q'', \ \mand  \ P'' \rl{R} Q'' \ )}$.

\smallskip\noindent
We denote with $\lbis$ the largest labelled bisimulation.
\end{definition}

In reactive synchronous programming, a program is usually supposed to
read `input' signals at the beginning of each instant and to react
delivering `output' signals at the end of each instant. In particular,
a program that does not reach a suspension point cannot produce an
observable output signal.  For instance, if we run 
$\ol{s} \mid \Omega$ then the emission on the
signal $s$ should not be observable because the program never
suspends. Following this intuition, we comment on the conditions $(L1-4)$.

\Proofitem{(L1)} This condition is standard in the framework of a
{\em bisimulation} semantics. As in the asynchronous case, it 
exposes the branching structure of a system to the extent that
it distinguishes, {\em e.g.}, the program $(\ol{s}_{1} \isum
\ol{s}_{2})\isum \ol{s}_{3}$ from the program $\ol{s}_{1} \isum
(\ol{s}_{2}\isum \ol{s}_{3})$. We will comment on alternative
approaches at the end of this section.

\Proofitem{(L2)}
According to the intuition sketched above, the condition $(L2)$ requires
that an output of a program $P$ is observable only if $P\lsusp$, {\em
i.e.}, only if $P$ may potentially reach a suspension point (remember
that in $\spi$ an output {\em persists} within an instant). The
reasons for choosing the L-suspension predicate rather than, {\em
e.g.}, the weak suspension predicate will be clarified in section
\ref{lab-suspension} and have to do with the fact that L-suspension
has better properties with respect to parallel composition. 
We also anticipate that in the premise of condition $(L2)$, 
it is equivalent to require $P\lsusp$ or $P'\lsusp$ 
(cf. remark \ref{alt-definition}) and that in the conclusion
the property $Q'\lsusp$ can be derived (cf. proposition
\ref{lsusp-lbis}).
Last but not least, we should stress that in practice we are
interested in programs that {\em react} at each instant and for this reason,
programs that do not satisfy the L-suspension predicate are usually
rejected by means of static analyses. In this relevant case, the condition
$(L2)$ is the usual output condition of the $\pi$-calculus.

\Proofitem{(L3)}
The reception of a signal is not directly observable just as the
reception of a message in the $\pi$-calculus with asynchronous
communication. For instance, there is no reason to distinguish $s.0,0$
from $0$. Techniques for handling this situation have already been
developed in the framework of the $\pi$-calculus with
{\em asynchronous communication} 
and amount to modify the input clause as in condition $(L3)$
(see \cite{ACS98}). It is a pleasant surprise that this idea can be
transposed to the current context.

\Proofitem{(L4)}
The condition $(L4)$ corresponds to the
end of the instant and of course it does not arise in the
$\pi$-calculus.  The end of the instant is an observable event since,
as we explained above, it is at the end of the instant that we 
get the results of the program for the current instant.
Let us explain the role of the context 
$S=\emit{s_{1}}{v_{1}} \mid \cdots \mid \emit{s_{n}}{v_{n}}$ 
in this condition. Consider the programs:
\[
P= s_1.0, A(!s_2)\qquad
Q= s_1.0,A([ ]) \qquad
A(l) = \matchv{l}{[ ]}{0}{\ol{s_{3}}}
\]
Then $P\susp$, $Q\susp$, 
$P\mapsto A([])$, and $Q\mapsto A([])$.
However, if we plug $P$ and $Q$ in the context 
$[\cdot ] \mid \ol{s}_2$
then the resulting programs exhibit different behaviours.
In other terms, when comparing two suspended programs we
should also consider the effect that emitted values
may have on the computation performed at the end of the instant.
We stress that the context $S$ must preserve the suspension of the program,
therefore the emissions in $S$ are only relevant if they
correspond to a signal $s$ which is dereferenced at the end of the
instant. In particular, the number of contexts $S$ to be considered in
rule $(L4)$ is finite whenever the number of distinct values that can be emitted
on dereferenced signals is finite (possibly up to injective renaming).
\\

Admittedly, the definition of labelled bisimulation is technical
and following previous work \cite{HY95,ACS98,FG98}, we seek its justification 
through suitable notions of  barbed and contextual bisimulation.  

\begin{definition}[commitment]\label{commit-def}
We write $P\commits \ol{s}$ if
$P\act{\outact{t}{s}{v}}\cdot$ and say that
$P$ commits to emit on $s$.
\end{definition}

\begin{definition}[barbed bisimulation]\label{barbed-bisimulation}
A symmetric relation $\rel{R}$ on programs is a barbed bisimulation if
whenever $P\rl{R} Q$ the following holds:

\Defitem{(B1)}
If $P\act{\tau} P'$ then
$\xst{Q'}{Q\wact{\tau} Q' \mand P'\rl{R}Q'}$.

\Defitem{(B2)}
If $P\commits \ol{s}$ and $P\lsusp$ then
$\xst{Q'}{(Q\wact{\tau} Q', Q'\commits \ol{s}, 
\mand P \rl{R}Q')}$.

\Defitem{(B3)}
If $P\susp$  and $P\mapsto P''$ then 
$\xst{Q', Q''}{(Q\wact{\tau} Q', Q'\susp, P \rl{R}Q',
Q'\mapsto Q'',  \mand 
P'' \rl{R}  Q'')}$.

\smallskip\noindent
We denote with $\bbis$ the largest barbed bisimulation.
\end{definition}

We claim that this is a `natural' definition.
Condition $(B1)$ corresponds to the usual treatment of $\tau$ moves.
Condition $(B2)$ corresponds to the observation of the output
commitments in the $\pi$-calculus with asynchronous communication
modulo the $L$-suspension predicate whose role has already
been discussed in presenting the condition $(L2)$.
We will see that the L-suspension predicate $\lsusp$ can be defined just in terms
of internal reduction (remark \ref{lbis-rmk}).
As in condition $(L2)$, the condition $Q'\lsusp$ is a
consequence of the definition (cf. proposition \ref{lsusp-cbis}(2)).
Finally, condition $(B3)$ corresponds to the observation of the
end of the instant and it is a special case of condition 
$(L4)$ where the context $S$ is empty.

\begin{definition} \label{static-cxt-def}
A static context $C$ is defined as follows:
\begin{equation}\label{contexts}
C::= [~] \Alt C\mid P \Alt \new{s}{C}
\end{equation}
\end{definition}

A reasonable notion of program equivalence should be preserved by the
static contexts, {\em i.e.}, by parallel composition and name generation. 
We define accordingly a notion of contextual bisimulation
(cf. \cite{HY95,FG98}).

\begin{definition}[contextual bisimulation]
A symmetric relation $\rel{R}$ on programs is a contextual bisimulation if
it is a barbed bisimulation (conditions $(B1-3)$) and moreover 
whenever $P\rl{R} Q$ then

\Defitem{(C1)} $C[P] \rl{R} C[Q]$, for any static context $C$.

\smallskip\noindent
We denote with $\cbis$ the largest contextual barbed bisimulation.
\end{definition}

Our main result shows that labelled and contextual
bisimulation collapse. In particular, this implies that labelled
bisimulation is preserved by the contexts $C$. The proof 
will be developed in the following sections.

\begin{theorem}\label{main-theorem}
Let $P,Q$ be programs. Then $P\lbis Q$ if and only if $P\cbis Q$.
\end{theorem}

We claim that our approach to the semantics of the $\spi$-calculus is
rather natural and mathematically robust, however we {\em cannot}
claim that it is more {\em canonical} than, say, the {\em weak, early
bisimulation semantics} of the $\pi$-calculus. 
We have chosen to explore a path following our mathematical taste,
however, as in the $\pi$-calculus, other paths could be
explored. In this respect, we will just mention 
three directions.  First, one could
remark that condition (B1) in definition \ref{barbed-bisimulation}
allows to observe the branching structure of a program and argue that
only suspended programs should be observed. This would lead us towards
a failure semantics/testing scenario \cite{CH98,BDP02} (in the testing
semantics, a program that cannot perform internal reductions is
called {\em stable} and this is similar to a {\em suspended} program
in the synchronous context).
Second, one could require that program equivalence is preserved by all
contexts and not just the static ones and proceed to adapt, say, the concept
of {\em open bisimulation} \cite{S96} to the present language.  
Third, one could plead for reduction congruence \cite{MS92} 
rather than for contextual
bisimulation and then try to see whether the two concepts coincide
following \cite{FG98}. We refer to the literature 
for standard arguments concerning bisimulation vs. testing semantics 
({\em e.g.}, \cite{Milner89}), 
early vs. open bisimulation ({\em e.g.}, \cite{S96}),
and contextual vs. reduction bisimulation ({\em e.g.}, \cite{FG98}).

\section{Understanding L-suspension}\label{lab-suspension}
In this section, we study the properties of the L-suspension predicate
and justify its use in the definition of labelled bisimulation.

\begin{proposition}[characterisations of L-suspension]\label{char-lsusp-prop}
Let $P$ be a program. The following are equivalent:

\Defitem{(1)} $P \lsusp$.

\Defitem{(2)} There is a program $Q$ such that $(P\mid Q)\wsusp$.

\Defitem{(3)} There is a static context $C$ (cf. definition \ref{static-cxt-def}) such that
$C[P] \lsusp$.
\end{proposition}
\Proof
\Proofitemf{(1\Arrow 2)}
Suppose $P_0 \act{\alpha_{1}} P_1 \cdots \act{\alpha_{n}} P_n$ and
$P_n\susp$.
We build $Q$ by induction on $n$.
If $n=0$ we can take $Q=0$. Otherwise, suppose $n>0$.
By inductive hypothesis, there is $Q_1$ such that
$(P_1 \mid Q_1)\wsusp$. We proceed by case analysis on the
first action $\alpha_1$.

\Defitem{(\alpha_1=\tau)} Then we can take $Q=Q_1$ and
$(P_0\mid Q)\act{\tau} (P_1 \mid Q_1)$.

\Defitem{(\alpha_1=sv)} 
Let $Q= (Q_1 \mid \emit{s}{v})$. We have
$(P_0\mid Q) \act{\tau} (P_1 \mid Q_1 \mid \emit{s}{v})$.
Since $P_1 \act{\ol{s}v} P_1$, 
we observe that $(P_1\mid Q_1) \wsusp$ implies
$(P_1 \mid Q_1 \mid \emit{s}{v}) \wsusp$.

\Defitem{(\alpha_1=\outact{t}{s}{v})}
We distinguish three subcases. 
\begin{enumerate}

\item If $\alpha_1=\emit{s}{t}$ then 
define $Q=s(t).Q_1$ and observe
that $(P_0\mid Q) \act{\tau} (P_1\mid Q_1)$.

\item If $\alpha_1=\new{t}{\emit{s}{t}}$
then define again $Q=s(t).Q_1$ and observe that
(i) $(P_0\mid Q) \act{\tau} \new{t}{(P_1\mid Q_1)}$ and
(ii)  $(P_1 \mid Q_1) \wsusp$ implies 
$\new{t}{(P_1\mid Q_1)}\wsusp$.

\item If $\alpha_1=\new{\vc{t}}{\emit{s}{\s{c}(\vc{v})}}$ then
let $\set{\vc{t'}}= \w{fn}(\s{c}(\vc{v}))\minus \set{\vc{t}}$
and $\vc{t''}$ a tuple of fresh names (one for each name in $\vc{t'}$).
We define 
$Q=s(x).\matchv{x}{[\vc{t''}/\vc{t'}]\s{c}(\vc{v})}{Q_1}{0}$ 
where $x,\vc{t''}\notin \w{FV}(Q_1)$ and observe that:
(i) $(P_0\mid Q) \wact{\tau} \new{\vc{t}}{(P_1\mid Q_1)}$ and
(ii)  $(P_1 \mid Q_1) \wsusp$ implies 
$\new{\vc{t}}{(P_1\mid Q_1)}\wsusp$. For instance,
if $P_0 \act{\new{t}{\emit{s}{\s{c}(t,t')}}} P_1$ then
we take $Q=s(x).\matchv{x}{\s{c}(t,t'')}{Q_1}{0}$ 
with $x,t''\notin\w{FV}(Q_1)$.

\end{enumerate}

\Proofitem{(2\Arrow 3)}
Take $C=[~]\mid Q$ and note that  by definition
$(P\mid Q)\wsusp$ implies $(P\mid Q)\lsusp$.

\Proofitem{(3\Arrow 1)}
First, check by induction on a static context $C$
that $P\act{\tau}\cdot$ implies
$C[P]\act{\tau} \cdot$. Hence, $C[P]\susp$ implies
$P\susp$.
Second, show that $C[P]\act{\alpha} Q$ implies
that $Q=C'[P']$ and either $P=P'$ or  $P\act{\alpha'}P'$.
Third, suppose 
$C[P] \act{\alpha_{1}} Q_1 \cdots \act{\alpha_{n}} Q_n$
with $Q_n\susp$. Show by induction on $n$ that
$P\lsusp$. \qed

\begin{remark}\label{lbis-rmk}
The second characterisation, shows that the
L-suspension predicate can be defined just in terms of the 
internal ($\tau$) transitions and the suspension predicate. 
Thus it does not depend on the choice of observing certain 
labels.
\end{remark}

\begin{proposition}[L-suspension and labelled equivalence]\label{lsusp-lbis}
\Defitemf{(1)}
If $\neg P \lsusp$ and $\neg Q \lsusp$ then
$P\lbis Q$.

\Defitem{(2)}
If $P \lbis Q$ and $P\lsusp$ then $Q\lsusp$.
\end{proposition}
\Proof
\Proofitemf{(1)}
First we note that $\neg P\lsusp$ and $P\act{\alpha}P'$
implies $\neg P'\lsusp$. Second, we check
that $R=\set{(P,Q) \mid \neg P \lsusp \mand \neg Q \lsusp}$
is a labelled bisimulation.

\Defitem{(L1)}
If $P\act{\tau} P'$ then $\neg P'\lsusp$. Then
$Q\wact{\tau} Q$ and $P'\rl{R} Q$.

\Defitem{(L2)}
The condition holds since $\neg P \lsusp$.

\Defitem{(L3)}
If $P\act{sv} P'$ then $\neg P'\lsusp$.
Then $Q\wact{\tau} Q$ and by proposition \ref{char-lsusp-prop},
$\neg Q \lsusp$ implies $\neg (Q\mid \emit{s}{v})\lsusp$.

\Defitem{(L4)} The condition holds since $\neg (P\mid S)\susp$.
Indeed if $(P\mid S)\susp$ then 
$(P\mid S)\lsusp$ and  by proposition  \ref{char-lsusp-prop}, 
$P\lsusp$ which contradicts the hypothesis. 

\Proofitem{(2)}
Suppose $P_0 \lbis Q_0$ and $P_0\lsusp$. 
We proceed by induction on the length $n$ of the shortest 
sequence of transitions to a suspended program:
$P_0 \act{\alpha_{1}} \cdots \act{\alpha_{n}} P_n$ and 
$P_n\susp$.
If $n=0$ then by $(L4)$, $Q_0\wact{\tau} Q'$ and $Q'\susp$. Thus
$Q_0\lsusp$.
If $n>0$ then we analyse the first action $\alpha_{1}$.

\Defitem{(\alpha_1=\tau)}
By $(L1)$, $Q_0\wact{\tau} Q_1$ and $P_1\lbis Q_1$. By
inductive hypothesis $Q_1\lsusp$ and therefore $Q_0\lsusp$.

\Defitem{(\alpha_1=\outact{t}{s}{v})}
By $(L2)$, since $P_0\lsusp$, we have $Q_0 \wact{\new{\vc{t}}{\emit{s}{v}}} Q_1$ and 
$P_1\lbis Q_1$. By inductive hypothesis, $Q_1\lsusp$. Thus $Q_0\lsusp$.

\Defitem{(\alpha_1=sv)} According to $(L3)$ we have two subcases.
If $Q_0\wact{sv} Q_1$ and $P_1\lbis Q_1$ then we reason
as in the previous case. 
If $Q_0\wact{\tau} Q_1$ and $P_1 \lbis (Q_1\mid \emit{s}{v})$ then
by inductive hypothesis $(Q_1\mid \emit{s}{v})\lsusp$.
By proposition \ref{char-lsusp-prop}, if 
$(Q_1\mid \emit{s}{v}) \lsusp$ then $Q_1\lsusp$. Thus $Q_0\lsusp$. \qed \\

Thus labelled bisimulation equates all programs which cannot L-suspend
and moreover it never equates a program which L-suspends to one which
cannot.  In this sense, L-suspension is reminiscent of the notion of
{\em solvability} in the $\lambda$-calculus 
\cite[p. 41]{B84}.  In spite of these nice
properties, one may wonder whether the L-suspension predicate could be
replaced by the suspension or weak suspension predicate.

\begin{definition}
We denote with $\lbissusp$ ($\lbiswsusp$) 
the notion of labelled bisimulation obtained by replacing 
in $(L2)$  the condition $P\lsusp$ with
the condition $P\susp$ ($P\wsusp$).
Similarly, we denote with $\bbissusp, \cbissusp$ 
($\bbiswsusp,\cbiswsusp$) 
the notions of barbed and contextual bisimulations obtained by replacing 
in $(B2)$  the condition $P\lsusp$ with 
the condition $P\susp$ ($P\wsusp$).
\end{definition}

\begin{proposition}[comparing bisimulations]
\Defitemf{(1)} The following inclusions hold:
\[
\bbis \ \subset\ \bbiswsusp\  \subset \ \bbissusp \ , 
\qquad
\lbis \ \subset \ \lbiswsusp \ \subset \ \lbissusp \ ,
\qquad
\cbis \ \subseteq \ \cbiswsusp \ \subseteq \ \cbissusp~.
\]
\Defitemf{(2)}
The barbed bisimulations 
and the labelled bisimulations $\lbiswsusp$ and 
$\lbissusp$ are not preserved by parallel composition.
\end{proposition}
\Proof
\Proofitemf{(1)}
The non-strict inclusions follow from the remark that
$P\susp$ implies $P\wsusp$ which implies $P\lsusp$.
We provide examples for the $4$ strict inclusions.

\Defitem{\bullet}
Consider $P=(\ol{s}_{1} \mid (\ol{s}_{2} \isum \ol{s}_{3}))$ and
$Q=(\ol{s}_{1} \mid \ol{s}_{2})\isum (\ol{s}_{1} \mid \ol{s}_{3})$.
Note that $P,Q\wsusp$ but $\neg P,Q \susp$ and that to reach a suspension
point, $P$ and $Q$ have to resolve their internal choices.
Now we have $P\lbissusp Q$ (and therefore $P\bbissusp Q$) but
$P\not\bbiswsusp Q$ (and therefore $P\not \lbiswsusp Q$).
To see the latter, observe that $P\commits \ol{s}_{1}$ and
that to match this commitment $Q$ must choose between $\ol{s}_{2}$ and
$\ol{s}_{3}$.

\Defitem{\bullet}
Let $(t,t')$ abbreviate $[t;t']$ and 
$s\arrow 0,\Omega$ abbreviate $s(x).\matchv{x}{\s{0}}{0}{\Omega}$.
Consider: 
\[
\begin{array}{ll}
P_1 &= \new{t,t'}{(\ol{s}(t,t') \mid (t.\ol{s}_{1} \isum t.\ol{s}_{2}) \mid Q) }\\

P_2 &= \new{t,t'}{ ( ( ( \ol{s}(t,t') \mid (t.\ol{s}_{1}) )\isum 
                       ( \ol{s}(t,t') \mid (t.\ol{s}_{2}) )) \mid Q) } \\

Q &= t'\arrow 0,\Omega \mid \ol{t'}{\s{1}}

\end{array}
\]
Note that $P_1,P_2 \lsusp$ but $\neg P_1,P_2 \wsusp$. 
The point is that the program $Q$ loops unless the name
$t'$ is extruded to the environment and the latter
provides a value $\s{0}$ on the signal $t'$.
Then $P_1\lbiswsusp P_2$. 
However,  $P_1\not\lbis P_2$.
To see this, notice that $P_1 \commits \ol{s}$
and that to match this commitment, $P_2$ has to resolve first the internal
choice between $\ol{s}_1$ and $\ol{s}_2$. A variant of this example where 
we remove the input prefix $t.\_$ before the emissions
$\ol{s}_{i}$,
$i=1,2$, shows that $\bbis$ is strictly included in $\bbiswsusp$.

\Proofitem{(2)}
It is well known that barbed bisimulation is not preserved by parallel
composition. For instance, $s.\ol{s}_{1} \bbis s.\ol{s}_{2}$, but
$(s.\ol{s}_{1} \mid \ol{s}) \not\bbis (s.\ol{s}_{2}\mid \ol{s})$ if
$s_1\neq s_2$.
To show that $\lbissusp$ and $\lbiswsusp$ are not preserved
by parallel composition consider again the programs $P_1$ and
$P_2$ above in parallel with:
\[
R = s(t,t').(
(\ol{t} \mid \ol{t'}\s{0}) 
\isum 
(\ol{t} \mid \ol{t'}\s{0} \mid \ol{s}_{3}))
\]
where $s(t,t').P$ abbreviates $s(x).\matchv{x}{[t;t']}{P}{0}$.
Remark that 
\[
(P_1\mid R) \wact{\tau} 
\new{t,t'}{(\ol{s}(t,t') \mid (t.\ol{s}_{1} \isum t.\ol{s}_{2}) \mid Q
  \mid \ol{t} \mid \ol{t'}\s{0})} \equiv P'_1
\]
To match this move, suppose $(P_2\mid R) \wact{\tau} P'_2$.
Now $P'_2$ must be able to suspend while losing the possibility of
committing on $\ol{s}_3$. Hence, there must be a synchronisation on
$s$ between $P_2$ and $R$. In turn, this synchronisation forces
$P_2$ to choose between $\ol{s}_1$ and $\ol{s}_2$.  Suppose, {\em e.g.},
$(P_2\mid R)$ chooses $\ol{s}_1$, then in a following move $P'_1$
chooses $\ol{s}_2$ and becomes:
\[
\new{t,t'}{(\ol{s}(t,t') \mid \ol{s}_{2} \mid 0 \mid  \ol{t} \mid
  \ol{t'}\s{0} \mid \ol{t'}\s{1})}
\]
which is suspended and commits on $\ol{s}_2$. 
The program $P'_2$ cannot match this move. \qed \\

Note that in (1) the inclusions for the barbed and labelled
bisimulations are strict. On the other hand, we do not know whether
the inclusions of the contextual bisimulations are strict.  However,
by (2) we do know that the notions of labelled bisimulation where
L-suspension is replaced by (weak) suspension are not preserved by
parallel composition and therefore cannot characterise the weaker
notions of contextual bisimulation. The conclusion we draw from this
analysis is that $\lbis$ is the good notion of labelled bisimulation
among those considered.

\section{Strong labelled bisimulation and an up-to technique}
It is technically convenient to introduce a {\em strong} notion of labelled
bisimulation which is used to bootstrap the reasoning about 
the weaker notion we are aiming at.

\begin{definition}[strong labelled bisimulation]\label{slb-def}
A symmetric relation $\rel{R}$ on programs is a strong 
labelled bisimulation if whenever $P\rl{R} Q$ the following holds:

\Defitem{(S1)} $P\act{\alpha} P'$ and $\w{bn}(\alpha)\inter
\w{fn}(Q)=\emptyset$ implies
$\xst{Q'}{(Q\act{\alpha} Q' \mand P'\rl{R} Q')}$.

\Defitem{(S2)} $(P\mid S) \susp$ with 
$S=(\emit{s_{1}}{v_{1}} \mid \cdots \mid \emit{s_{n}}{v_{n}})$, $n\geq 0$ 
and $(P\mid S) \mapsto P'$ implies 
$(P \mid S) \rl{R} (Q \mid S)$ and  $\xst{Q'}{(Q\mapsto Q' \mand
P' \rl{R} Q')}$.

\smallskip\noindent
We denote with $\sbis$ the largest strong labelled bisimulation.
\end{definition}

\begin{proposition}\label{strong-label}
If $P\sbis Q$ then $P\lbis Q$.
\end{proposition}
\Proof We check that $\sbis$ is a labelled bisimulation.
Conditions $(L1-3)$ follow from condition $(S1)$.
Condition $(L4)$ follows from condition $(S2)$ noticing that
$(P \mid S) \sbis (Q \mid S)$ and $(P\mid S)\dcl$ implies by $(S1)$
that $(Q\mid S)\susp$. \qed \\

When comparing strong labelled bisimulation with labelled bisimulation
it should be noticed that in the former not only we forbid
weak internal moves but we also drop the convergence condition
in $(L2)$ and the possibility of matching an input with
an internal transition in $(L3)$. For this reason, we adopt
the notation $\sbis$ rather than the usual $\sim_L$.

\begin{definition}
We say that a relation $\rel{R}$ is a strong labelled bisimulation up to
strong labelled bisimulation if the conditions $(S1-2)$ hold when
we replace $\rel{R}$ with the larger relation $(\sbis) \comp \rel{R} \comp
(\sbis)$.
\end{definition}

The following proposition summarizes some 
useful properties of strong labelled bisimulation.
In the present context, an {\em injective renaming} is an
injective function mapping signal names to signal names.

\begin{proposition}[properties of $\sbis$]\label{sbis-prop}
\Defitemf{(1)}
If $P\sbis Q$ and $\sigma$ is an injective renaming then
$\sigma P \sbis \sigma Q$.

\Defitem{(2)}
$\sbis$ is a reflexive and transitive relation.

\Defitem{(3)} The following laws hold:
\[
\begin{array}{c}
(P\mid 0) \sbis P,
\qquad
P_1 \mid (P_2 \mid P_3) \sbis (P_1\mid P_2) \mid P_3, 
\qquad (P_1 \mid P_2) \sbis (P_2 \mid P_1), \\
\new{s_1,s_2}{P} \sbis \new{s_2,s_1}{P}
\qquad
\new{s}{P_1} \mid P_2 \sbis \new{s}{(P_1 \mid P_2)}
\mbox{ if }s\notin \w{fn}(P_2).
\end{array}
\]
\Defitemf{(4)}
If $P\sbis Q$ then $(P\mid S) \sbis (Q\mid S)$ where
$S=(P_1 \mid \cdots \mid  P_n)$ and 
$P_i=0$ or $P_i=\emit{s_{i}}{v_{i}}$, for $i=1,\ldots,n$, $n\geq 0$.

\end{proposition}
\Proofhint 
Most properties follow by routine verifications.
We just highlight some points.

\Proofitem{(2)}
Recalling that $P\sbis Q$ and $P\susp$ implies $Q\susp$.

\Proofitem{(3)}
Introduce a notion of normalised program where 
parallel composition associates to the left,
all restrictions are carried at top level, 
and $0$ programs are the identity for parallel composition.
Then define a relation $\rel{R}$ where two programs are
related if their normalised forms are identical
up to bijective permutations of the restricted names
and the parallel components. A pair of programs equated by the laws
under consideration is in $\rel{R}$.
Show that $\rel{R}$ is a strong labelled bisimulation. 

\Proofitem{(4)}
Show that 
$\set{(P\mid S, Q\mid S) \mid P\sbis Q}$ is a strong
labelled bisimulation where $S$ is defined 
as in the statement. \qed  \\

The following proposition summarizes the properties of
the output transition.

\begin{proposition}[emission]\label{emission-prop}

\Defitemf{(1)}
If $P\act{\outact{t}{s}{v}} P'$ then 
$P \sbis \new{\vc{t}}{(\ol{s}v \mid P'')}$ and 
$P'\sbis (\ol{s}v \mid P'')$.

\Defitem{(2)}
If $P\act{\outact{t}{s}{v}} P'$ then $P\lsusp$ if and only if
$P'\lsusp$. 
\end{proposition}
\Proof  
\Proofitemf{(1)}
In deriving $P\act{\outact{t}{s}{v}} P'$ one can only rely on the rules
$(\w{out},\w{par},\nu,\nu_{\w{ex}})$. We use the
laws of strong labelled bisimulation (proposition \ref{sbis-prop}(2)) to put
the program in the desired form.

\Proofitem{(2)} 
By definition, $P'\lsusp$ implies $P\lsusp$. In the other direction,
relying on  (1), assume that the program has the shape 
$\new{\vc{t}}{(\emit{s}{v} \mid P)}$. We also know that this program
L-suspends. By proposition \ref{char-lsusp-prop}, there is a program $Q$ such
$\new{\vc{t}}{(\emit{s}{v} \mid P)}\mid Q \wsusp$. That is, assuming
$\set{\vc{t}}\inter \w{fn}(Q)=\emptyset$, we have that
$\new{\vc{t}}{(\emit{s}{v} \mid P \mid Q)} \wsusp$.
The latter implies that there is a $Q'$ such that
$(\emit{s}{v} \mid P \mid Q) \wact{\tau} Q'$ and $Q'\susp$.
Again, by proposition \ref{char-lsusp-prop}, this means that 
$(\emit{s}{v}\mid P)\lsusp$. \qed

\begin{remark}\label{alt-definition}
By proposition \ref{emission-prop}(2), 
in condition $(L2)$ of definition \ref{lab-bis-def},
it is equivalent to  require $P\lsusp$ or $P'\lsusp$.
\end{remark}

Our main application of strong labelled bisimulation is in the context of 
a rather standard `up to technique'.

\begin{definition}
A relation $\rel{R}$ is a labelled bisimulation up to $\sbis$ if 
the conditions $(L1-4)$ are satisfied when we
replace the relation $\rel{R}$ with the (larger) relation 
$(\sbis)\comp \rel{R} \comp (\sbis)$.
\end{definition}

\begin{proposition}[up-to technique]
Let $\rel{R}$ be a labelled bisimulation up to $\sbis$. Then:

\Defitem{(1)} 
The relation $(\sbis)\comp \rel{R} \comp (\sbis)$ is a labelled
bisimulation.

\Defitem{(2)} If $P\rl{R} Q$ then $P\lbis Q$.
\end{proposition}
\Proof
\Proofitemf{(1)}
A direct diagram chasing using proposition \ref{sbis-prop}.

\Proofitem{(2)}
Follows directly from (1). \qed

\section{Congruence properties of labelled bisimulation}
We are now ready to study the congruence properties of 
labelled bisimulation. The most important part of the 
proof concerns the preservation under parallel composition
and name generation and it is composed of $12$ cases.

\begin{proposition}\label{prop-lbis-congr}

\Defitem{(1)} If $P_1\lbis P_2$ and $\sigma$ is an injective renaming
then $\sigma P_1 \lbis \sigma P_2$.

\Defitem{(2)}
If $P_1 \lbis P_2$ then $(P_1 \mid \emit{s}{v}) \lbis (P_2 \mid \emit{s}{v})$.

\Defitem{(3)}
The relation $\lbis$ is reflexive and transitive.

\Defitem{(4)} If $P_1\lbis P_2$ then $\new{s}{P_{1}} \lbis \new{s}{P_{2}}$ and
$(P_1\mid Q) \lbis (P_2 \mid Q)$.

\end{proposition}
\Proof 
\Proofitemfb{(1)} By propositions \ref{sbis-prop}(1) and \ref{strong-label}.

\Proofitemb{(2)} 
We show that the relation 
$\rel{R}=\lbis \union 
\set{( \ P_1 \mid \emit{s}{v}, P_2 \mid \emit{s}{v} \  ) \mid P_1 \lbis  P_2}$
is a labelled bisimulation up to $\sbis$. 
We assume $P_1 \lbis P_2$ and we analyse the conditions $(L1-4)$.

\Proofitem{(L1)} Suppose
$(P_1 \mid \emit{s}{v}) \act{\tau} (P'_{1} \mid \emit{s}{v})$. If the
action $\tau$ is performed by $P_1$ then the hypothesis and condition
$(L1)$ allow to conclude. Otherwise, suppose $P_1 \act{sv} P'_{1}$. Then we
apply the hypothesis and condition $(L3)$. Two cases may arise:
(1) If $P_2 \wact{sv} P'_{2}$ and $P'_{1} \lbis P'_{2}$ then the conclusion is
immediate.
(2) If $P_2 \wact{\tau} P'_{2}$ and $P'_{1} \lbis (P'_{2} \mid \emit{s}{v})$ then
we note that $(P'_{2}\mid \emit{s}{v}) \sbis (P'_{2} \mid \emit{s}{v}) \mid 
\emit{s}{v}$ and we close the diagram up to $\sbis$.

\Proofitem{(L2)} Suppose 
$(P_1 \mid \emit{s}{v})\lsusp$ and 
$(P_1\mid \emit{s}{v}) \act{\outact{t}{s'}{v}} (P'_1 \mid \emit{s}{v})$.
If the emission action is performed by $\emit{s}{v}$ then the conclusion
is immediate.
Otherwise, note that $P_1\lsusp$. Hence by $(L2)$,
$P_2 \wact{\outact{t}{s'}{v}} P'_2$ and $P'_1 \lbis P'_2$.
But then $(P_2\mid \emit{s}{v}) \wact{\outact{t}{s'}{v}} (P'_2\mid \emit{s}{v})$
and we can conclude.

\Proofitem{(L3)} Suppose 
$(P_1\mid \emit{s}{v})\act{s'v'} (P'_1\mid \emit{s}{v})$.
Necessarily, $P_1\act{s'v'} P'_1$. By $(L3)$, two cases may arise.
If $P_2\wact{s'v'}P'_2$ and $P'_1 \lbis P'_2$ then the conclusion
is direct.
On the other hand, if $P_2\wact{\tau} P'_2$ and $P'_1 \lbis (P'_2 \mid \emit{s'}{v'})$
then we note that 
\[
(P'_1 \mid \emit{s}{v}) \rl{R} ((P'_2 \mid \emit{s'}{v'})\mid
\emit{s}{v}) 
\sbis ((P'_2 \mid \emit{s}{v})\mid \emit{s'}{v'})
\]
and we close the diagram up to $\sbis$.

\Proofitem{(L4)}
Let $S=\emit{s_{1}}{v_{1}} \mid \cdots \mid \emit{s_{n}}{v_{n}}$.
Suppose $(P_1 \mid \emit{s}{v} \mid S) \susp$ and $(P_1 \mid
\emit{s}{v}\mid S) \mapsto P'_1$.
By $(L4)$ applied to $(\emit{s}{v} \mid S)$, we derive that 
$(P_2\mid \emit{s}{v} \mid S) \wact{\tau} (P''_{2} \mid  \emit{s}{v} \mid
S)$, $(P''_{2} \mid  \emit{s}{v} \mid S) \susp$, 
$(P_1 \mid \emit{s}{v} \mid S) \lbis (P''_{2} \mid  \emit{s}{v} \mid S)$,
$(P''_{2} \mid  \emit{s}{v} \mid S) \mapsto P'_{2}$, and
$P'_1 \lbis P'_2$.

\Proofitemb{(3)} 
It is easily checked that the identity relation is a labelled
bisimulation. Reflexivity follows. 
As for transitivity, we check that the relation 
$R= \lbis \comp \lbis$ is a labelled bisimulation up to $\sbis$.
Suppose $P_1\lbis P_2 \lbis P_3$.

\Proofitem{(L1)}  Standard argument.

\Proofitem{(L2)} Suppose $P_1 \lsusp$ and 
$P_1 \act{\outact{t}{s}{v}} P'_1$.
Note that by (1) we can assume that the names $\vc{t}$ are not in $P_2$.
By $(L2)$, $P_2 \wact{\outact{t}{s}{v}} P'_2$ and 
$P'_1 \lbis P'_2$. By proposition \ref{emission-prop}(2), $P_1 \lsusp$ implies
$P'_1 \lsusp$. By proposition \ref{lsusp-lbis}(2),
$P'_1 \lsusp$ and  $P'_1 \lbis P'_2$ implies
$P'_2 \lsusp$.
We conclude by applying $(L1)$ and $(L2)$ to $P_2$ and $P_3$.

\Proofitem{(L3)} 
Suppose $P_1 \act{sv} P'_1$. Two interesting cases arise when either
$P_2$ or $P_3$ match an input action with an internal transition.
(1) Suppose first $P_2 \wact{\tau} P'_2$ and 
$P_1 \lbis (P'_2 \mid \emit{s}{v})$. By $P_2\lbis P_3$ and
repeated application of $(L1)$ we derive that
$P_3 \wact{\tau} P'_3$ and $P'_2 \lbis P'_3$. 
By property (2), the latter implies that
$(P'_2 \mid \emit{s}{v}) \lbis (P'_3 \mid \emit{s}{v})$ and
we combine with $P_1 \lbis (P'_2 \mid \emit{s}{v})$ to conclude.
(2) Next suppose 
$P_2 \wact{\tau} P^{1}_{2} \act{sv} P^{2}_{2} \wact{\tau} P'_2$ and
$P_1 \lbis P'_2$. Suppose that $P_3$ matches these transitions as
follows:
$P_3 \wact{\tau} P^{1}_{3} \wact{\tau} P^{2}_{3}$, 
$P^{2}_{2}\lbis (P^{2}_{3} \mid \emit{s}{v})$, and moreover
$(P^{2}_{3} \mid \emit{s}{v}) \wact{\tau} (P'_3  \mid \emit{s}{v})$ with
$P'_2 \lbis (P'_3\mid \emit{s}{v})$. 
Two subcases may arise:
(i) $P^{2}_{3} \wact{\tau}  P'_3$. Then we have 
$P_3 \wact{\tau} P'_3$, $P'_2 \lbis  (P'_3\mid \emit{s}{v})$ and
we can conclude.
(ii) $P^{2}_{3} \wact{sv}  P'_3$.
Then we have $P_3 \wact{sv} P'_3$ and 
$P'_2 \lbis (P'_3 \mid \emit{s}{v}) \sbis P'_3$.
Note that $P^{2}_{3}$ does not need to perform the action
$sv$ more than once.

\Proofitem{(L4)}
Let $S=\emit{s_{1}}{v_{1}} \mid \cdots \mid \emit{s_{n}}{v_{n}}$.
Suppose $(P_1\mid S)\susp$ and $(P_1\mid S)\mapsto P'_1$.
By $(L4)$, $(P_2\mid S) \wact{\tau} (P''_2 \mid S)$,
$(P''_2 \mid S)\susp$, $(P_1\mid S)\lbis (P''_2 \mid S)$,
$(P''_2 \mid S)\mapsto P'_2$, and $P'_1 \lbis P'_2$.
By $(L1)$, $(P_3\mid S) \wact{\tau} (P''_3 \mid S)$ and 
$(P''_2 \mid S) \lbis (P''_3 \mid S)$.
By $(L4)$, $(P''_3 \mid S)\wact{\tau} (P'''_3 \mid S)$,
$(P'''_3 \mid S)\susp$, $(P''_2 \mid S) \lbis (P'''_3 \mid S)$,
$(P'''_3 \mid S)\mapsto P'_3$, $P'_2 \lbis P'_3$ and we can conclude.

\Proofitemb{(4)} 
We show that  
$\rel{R}=\set{(\new{\vc{t}}{(P_1 \mid Q)}, \new{\vc{t}}{(P_2\mid Q)})  \mid
P_1 \lbis P_2} \union \lbis$ is a  labelled bisimulation up to $\sbis$.

\Proofitem{(L1)} Suppose $\new{\vc{t}}{(P_1 \mid Q)} \act{\tau}\cdot$.
This may happen because either $P_1$ or $Q$ perform a $\tau$ action
or because $P_1$ and $Q$ synchronise. We consider the various
situations that may occur.

\Defitem{(L1)[1]} Suppose $Q\act{\tau} Q'$. Then
$\new{\vc{t}}{(P_2 \mid Q)} \act{\tau}\new{\vc{t}}{(P_2 \mid Q')}$
and we can conclude.

\Defitem{(L1)[2]} Suppose $P_1 \act{\tau} P'_1$. By $(L2)$ 
$P_2\wact{\tau} P'_2$ and $P'_1 \lbis P'_2$.
Then $\new{\vc{t}}{(P_2 \mid Q)} \wact{\tau}\new{\vc{t}}{(P'_2 \mid
  Q)}$  and we can conclude.

\Defitem{(L1)[3]} Suppose 
$P_1 \act{sv} P'_1$ and $Q\act{\outact{t'}{s}{v}}Q'$.
According to $(L3)$, we have two subcases.

\Defitem{(L1)[3.1]} Suppose  $P_2\wact{sv} P'_2$ and $P'_1\lbis P'_2$.
Then  $\new{\vc{t}}{(P_2 \mid Q)} \wact{\tau}\new{\vc{t},\vc{t'}}{(P'_2 \mid
  Q')}$ and we can conclude.

\Defitem{(L1)[3.2]}  Suppose $P_2\wact{\tau} P'_2$ and 
$P'_1\lbis (P'_2\mid \emit{s}{v})$. 
By proposition \ref{emission-prop}(2), $Q \sbis \new{\vc{t'}}{Q'}$ and 
$Q'\sbis (Q'' \mid \emit{s}{v})$ for some $Q''$.
Then 
 $\new{\vc{t}}{(P_2 \mid Q)} \wact{\tau} 
  \new{\vc{t}}{(P'_2 \mid Q)} \sbis 
  \new{\vc{t},\vc{t'}}{(P'_2 \mid \emit{s}{v})\mid Q''}$ 
and we can conclude up to $\sbis$.

\Defitem{(L1)[4]} Suppose $P_1 \act{\outact{t'}{s}{v}} P'_1$ and 
$Q\act{sv} Q'$. We have two subcases.

\Defitem{(L1)[4.1]} Suppose $\neg P_1 \lsusp$. By propositions
\ref{char-lsusp-prop} and \ref{lsusp-lbis},
$\neg \new{\vc{t}}{(P_1 \mid Q)} \lsusp$, $\neg P_2\lsusp$, 
$\neg \new{\vc{t}}{(P_2 \mid Q)}\lsusp$, 
$\neg P'_1 \lsusp$, and $\neg  \new{\vc{t},\vc{t'}}{(P'_1 \mid Q')} \lsusp$.
Hence, $\new{\vc{t},\vc{t'}}{(P'_1 \mid Q')} \lbis 
\new{\vc{t}}{(P_2 \mid Q)}$ and we can conclude.

\Defitem{(L1)[4.2]} Suppose $P_1 \lsusp$. By $(L2)$, 
$P_2 \wact{\outact{t'}{s}{v}} P'_2$ and $P'_{1}\lbis P'_{2}$. Hence
$\new{\vc{t}}{(P_2 \mid Q)}\wact{\tau} \new{\vc{t},\vc{t'}}{(P'_2 \mid
  Q')}$ and we can conclude. 

\Proofitem{(L2)} Suppose 
$\new{\vc{t}}{(P_1 \mid Q)} \act{\outact{t'}{s}{v}} \cdot$ and 
$\new{\vc{t}}{(P_1 \mid Q)} \lsusp$.
Also assume
$\vc{t}=\vc{t_{1}},\vc{t_{2}}$ and
$\vc{t'} = \vc{t_{1}},\vc{t_{3}}$ up to reordering so that
the emission extrudes exactly the names $\vc{t}_{1}$ among the names in 
$\vc{t}$.
We have two subcases depending which component performs the action.

\Defitem{(L2)[1]} Suppose 
 $Q \act{\outact{t_{3}}{s}{v}} Q'$. 
Then $\new{\vc{t}}{(P_2 \mid Q)} \act{\outact{t'}{s}{v}} 
\new{\vc{t_{2}}}{(P_2 \mid Q')}$ and we can conclude.

\Defitem{(L2)[2]} Suppose $P_1 \act{\outact{t_{3}}{s}{v}} P'_1$.
By proposition \ref{char-lsusp-prop}, we know that $P_1\lsusp$. 
Hence $P_2 \wact{\outact{t_{3}}{s}{v}} P'_2$ and 
$P'_1 \lbis P'_2$. Then
$\new{\vc{t}}{(P_2 \mid Q)} \act{\outact{t'}{s}{v}} 
\new{\vc{t_{2}}}{(P'_2 \mid Q)}$ and we can conclude.

\Proofitem{(L3)} Suppose 
$\new{\vc{t}}{(P_1 \mid Q)} \act{sv} \cdot$ 
We have two subcases  depending which component performs the action.

\Defitem{(L3)[1]} Suppose  $Q\act{sv} Q'$. 
Then $\new{\vc{t}}{(P_2 \mid Q)} \act{sv} 
\new{\vc{t}}{(P_2 \mid Q')}$ and we can conclude.

\Defitem{(L3)[2]} Suppose $P_1 \act{sv} P'_1$. 
According to $(L3)$ we have two subcases.

\Defitem{(L3)[2.1]} Suppose 
$P_2 \wact{sv} P'_2$ and $P'_1 \lbis P'_2$.
Then $\new{\vc{t}}{(P_2 \mid Q)} \wact{sv} 
\new{\vc{t}}{(P'_2 \mid Q)}$ and we can conclude.

\Defitem{(L3)[2.2]} Suppose
$P_2 \wact{\tau} P'_2$ and $P'_1 \lbis (P'_2\mid \emit{s}{v})$.
Then $\new{\vc{t}}{(P_2 \mid Q)} \wact{\tau} 
\new{\vc{t}}{(P'_2 \mid Q)}$ and 
since $\new{\vc{t}}{(P'_2 \mid Q)} \mid \emit{s}{v} \sbis
\new{\vc{t}}{((P'_2 \mid \emit{s}{v}) \mid Q)}$ we
can conclude up to $\sbis$.

\Proofitem{(L4)} Suppose 
$S=\emit{s_{1}}{v_{1}} \mid \cdots \mid \emit{s_{n}}{v_{n}}$ and 
$\new{\vc{t}}{(P_1 \mid Q)} \mid S \susp$. 
Up to strong labelled bisimulation, we can express $Q$ as
$\new{\vc{t}_Q}{(S_Q \mid I_Q)}$ where $S_Q$ is the
parallel composition of emissions and $I_Q$ is the
parallel composition of receptions.
Thus we have:
$\new{\vc{t}}{(P_1 \mid Q)} \mid S \sbis
 \new{\vc{t},\vc{t}_Q}{(P_1 \mid S_Q \mid I_Q \mid S)}$,
and 
$\new{\vc{t}}{(P_2 \mid Q)} \mid S \sbis
 \new{\vc{t},\vc{t}_Q}{(P_2 \mid S_Q \mid I_Q \mid S)}$
assuming $\set{\vc{t}}\inter \w{fn}(S)=\emptyset$ and 
$\set{\vc{t}_{Q}}\inter \w{fn}(P_i \mid S)=\emptyset$
for $i=1,2$.

If $\new{\vc{t}}{(P_1 \mid Q)} \mid S \mapsto P$ then
$P\sbis \new{\vc{t},\vc{t}_Q}{(P''_1 \mid Q')}$ where in
particular, we have that
$(P_1\mid S_Q \mid S)\dcl$ and $(P_1 \mid S_Q \mid S) \mapsto 
(P'_1 \mid 0 \mid 0)$.

By the hypothesis $P_1\lbis P_2$ and $(L4)$ we derive that: 
(i) $(P_2 \mid S_Q \mid S) \wact{\tau} (P''_{2} \mid S_Q \mid S)$,
(ii) $(P''_{2} \mid S_Q \mid S)\susp$,
(iii) $(P''_{2} \mid S_Q \mid S)\mapsto (P'_{2} \mid 0 \mid 0)$,
(iv) $(P_1\mid S_Q \mid S)\lbis (P''_{2} \mid S_Q \mid S)$, and
(v) $(P'_1 \mid 0 \mid 0) \lbis (P'_{2} \mid 0 \mid 0)$.

Because $(P_1\mid S_Q \mid S)$ and $(P''_{2} \mid S_Q \mid S)$ are
suspended and labelled bisimilar, the two programs must
commit (cf. definition \ref{commit-def}) on the same signal names
and moreover on each signal name they must emit the same set of
values up to renaming of bound names. It follows that the program
$\new{\vc{t},\vc{t}_Q}{(P''_{2} \mid S_Q \mid I_Q \mid S)}$ is
suspended. The only possibility for an internal transition 
is that an emission in $P''_{2}$ enables a reception in $I_Q$ but
this contradicts the hypothesis that 
$\new{\vc{t},\vc{t}_Q}{(P_{1} \mid S_Q \mid I_Q \mid S)}$ 
is suspended.
Moreover, $(P''_{2} \mid S_Q \mid I_Q \mid S)\mapsto
(P'_{2} \mid 0 \mid Q' \mid 0)$.

Therefore, we have that 
\[
\new{\vc{t}}{(P_2 \mid Q)} \mid S \sbis
 \new{\vc{t},\vc{t}_Q}{(P_2 \mid S_Q \mid I_Q \mid S)}
 \wact{\tau}
 \new{\vc{t},\vc{t}_Q}{(P''_2 \mid S_Q \mid I_Q \mid S)},
\]
$\new{\vc{t},\vc{t}_Q}{(P''_2 \mid S_Q \mid I_Q \mid S)}\susp$, 
and 
$\new{\vc{t},\vc{t}_Q}{(P''_2 \mid S_Q \mid I_Q \mid S)} \mapsto 
 \new{\vc{t},\vc{t}_Q}{(P'_2 \mid 0 \mid Q' \mid 0)}$.
Now  $\new{\vc{t},\vc{t}_Q}{(P_1 \mid S_Q \mid I_Q \mid S)}
      \rl{R}
      \new{\vc{t},\vc{t}_Q}{(P''_2 \mid S_Q \mid I_Q \mid S)}$
because $(P_1 \mid S_Q  \mid S)\lbis (P''_2 \mid S_Q \mid S)$ and
$\new{\vc{t},\vc{t}_Q}{(P'_1 \mid Q')}
      \rl{R}
      \new{\vc{t},\vc{t}_Q}{(P'_2 \mid Q')}$
because $P'_1 \lbis P'_2$. \qed \\

We can now derive the first half of the proof of theorem
\ref{main-theorem}.

\begin{corollary}\label{lbis-impl-cbis}
Let $P,Q$ be programs. Then $P\lbis Q$ implies $P\cbis Q$.
\end{corollary}
\Proof Labelled bisimulation is a
barbed bisimulation and by proposition \ref{prop-lbis-congr} 
it is preserved by the contexts $C$. Hence it
is a contextual bisimulation. \qed

\section{Building discriminating contexts}
To complete the proof of theorem \ref{main-theorem}, 
it remains to show that our contexts are sufficiently
strong to make all distinctions labelled bisimulation
does. First we note the analogous of proposition \ref{lsusp-lbis} 
for contextual bisimulation.

\begin{proposition}\label{lsusp-cbis}
\Defitemf{(1)}
If $\neg P \lsusp$ and $\neg Q \lsusp$ then
$P\cbis Q$.

\Defitem{(2)}
If $P \cbis Q$ and $P\lsusp$ then $Q\lsusp$.
\end{proposition}
\Proof 
\Proofitemf{(1)} 
By proposition \ref{lsusp-lbis}, $P\lbis Q$ and 
by corollary \ref{lbis-impl-cbis}, $P\cbis Q$.

\Proofitem{(2)}
By proposition \ref{char-lsusp-prop}, there is 
a program $R$ such that 
$(P\mid R)\wsusp$, {\em i.e.},
$(P\mid R)\wact{\tau} P_1$ and $P_1 \susp$.
By $(C1)$, $(P\mid R)\cbis (Q\mid R)$.
By $(B1)$, $(Q\mid R) \wact{\tau} Q'_1$ and
$P_1\cbis Q'_1$. By $(B3)$, $Q'_1\wact{\tau} Q_1$ 
and $Q_1\susp$. Thus $(Q\mid R) \wsusp$ and
again by proposition \ref{char-lsusp-prop} this
implies that $Q\lsusp$. \qed

\begin{proposition}
If $P\cbis Q$ then $P\lbis Q$.
\end{proposition}
\Proof We denote with $a_i,b_i,c_i,\ldots$ `fresh' signal names 
not occurring in the programs under consideration.
We will rely on the signal names $a_i$ to extrude the scope of
some signal names and on the signal names $b_i,c_i$ 
to monitor the internal transitions of the programs.
We define a relation $\cl{R}$:
\[
\begin{array}{ll}
P_1 \rl{R} P_2  
&\mbox{if } \new{\vc{t}}{(P_1\mid O)} \cbis \new{\vc{t}}{(P_2\mid O)} 
\mbox{ for some }\vc{t},O,  \\
&\mbox{where: }
\vc{t}=t_1\ldots,t_n, 
O=\emit{a_{1}}{t_{1}} \mid \cdots \mid \emit{a_{n}}{t_{n}}, 
\set{a_1,\ldots,a_n}\inter\w{fn}(P_1\mid P_2)=\emptyset.
\end{array}
\]
By definition, if $P_1 \cbis P_2$ then $P_1 \rl{R} P_2$ taking
$\vc{t}$ as the empty vector and $O$ as the empty parallel
composition.  The purpose of the relation $\cl{R}$ is to enlarge the definition
of contextual bisimulation so that some signal names $\vc{t}$ are at once
restricted and observable thanks to the emission performed by $O$. 
We will will show that $\rl{R}$ is a labelled bisimulation up to strong 
labelled bisimulation so that we have the following implications:
\[
P_1 \cbis P_2  \quad \Arrow \quad P_1 \rl{R} P_2
               \quad \Arrow \quad P_1 \lbis P_2~.
\]

\Proofitem{\bullet}
We have seen in section \ref{derived-op} that an internal choice operator $\isum$
is definable in the $\spi$-calculus. In order to simplify the
notation, in the following we assume that 
$P_1 \isum P_2$ reduces to either $P_1$ or $P_2$ by just {\em one}
$\tau$-transition.  In reality, the reduction takes 
one $\tau$-transition to perform the internal choice, a second deterministic
$\tau$-transition to select the right branch of the matching operator,
and some garbage collection to remove signals that are under the scope
of a restriction and cannot be received.  The second transition and the garbage
collection do not affect the structure of the proof and we will ignore them.

\Proofitem{\bullet} 
Assuming $O=\emit{a_{1}}{t_{1}} \mid \cdots \mid \emit{a_{n}}{t_{n}}$
and $\vc{a} =a_1,\ldots,a_n$,
we will repeatedly use a program $R(\vc{a})[P]$ which is defined
as follows:
\[
\begin{array}{lll}
R(\vc{a})[P] &= &a_{1}(t_{1}). \ol{b_{1}} \isum (\ol{c_{1}} \isum  \\
             &   &\quad a_{2}(t_{2}). \ol{b_{2}} \isum (\ol{c_{2}} \isum  \\
             &   &\quad \quad \ldots \\
             &   &\quad \quad \quad a_{n}(t_{n}). \ol{b_{n}} \isum (\ol{c_{n}} \isum P)\ldots )
\end{array}
\]
Next we assume $P_1\rl{R} P_2$ 
because $\new{\vc{t}}{(P_1\mid O)} \cbis \new{\vc{t}}{(P_2\mid O)}$
for some $\vc{t},O$,  and consider the conditions $(L1-4)$.

\Proofitem{(L1)}
Suppose $P_1 \act{\tau} P'_1$.
Then $\new{\vc{t}}{(P_1\mid O)} \act{\tau} 
\new{\vc{t}}{(P'_1\mid O)}$.
By $(B1)$, $\new{\vc{t}}{(P_2\mid O)} \wact{\tau} Q$
and $\new{\vc{t}}{(P'_1\mid O)}\cbis Q$.
Note however that $O$ cannot interact with $P_2$ and its
derivatives because the signal names $\vc{a}$ do not
occur in $(P_1\mid P_2)$. Hence it must be that
$P_2 \wact{\tau} P'_2$ and $Q= \new{\vc{t}}{(P'_2\mid O)}$.
Then by definition of the relation $\cl{R}$, we derive that 
$P'_1 \rl{R} P'_2$.

\Proofitem{(L2)}
Suppose $P_1\lsusp$ and $P_1 \act{\outact{t'}{s}{v}} P'_1$ with
$\vc{t'}=t'_1,\ldots,t'_m$.
Let $X=\w{fn}(P_1\mid P_2)$.
Let 
\[
\begin{array}{ll}
R  & =R(\vc{a})[s(x).[x=\new{\vc{t'}}{v}]_{X\union \set{\vc{t'}}} \ (\ol{b_{n+1}} \isum (\ol{c_{n+1}}
      \isum O'))], \mbox{ where } \\
O' &=a_{n+1}t'_{1} \mid \cdots \mid a_{n+m}t'_{m}
\end{array}
\]
Now we have:
\[
\new{\vc{t}}{(P_1 \mid O)} \mid R \wact{\tau} \new{\vc{t},\vc{t'}}{(P'_1 \mid O \mid O')}
\]
by a series of reductions where first $R$ interacts with $O$ 
to learn the names $t_1\ldots,t_n$, then it 
interacts with $P_1$ to read a value $\new{\vc{t'}}{v}$ 
(note that the freshness of $\vc{t'}$ is checked with respect to both
$X$ and $\vc{t}$), and
finally it emits with $O'$ the names $\vc{t'}$ extruded by $P_1$.
We remark that in all the intermediate steps the program has the L-suspension
property, thus condition $(B2)$ applies and in particular the 
commitments on 
$\ol{b}_{i},\ol{c}_{i}$ are observable.

Next, we decompose this series of reductions in several steps and
analyse how the program $\new{\vc{t}}{(P_2 \mid O)} \mid R$ may match
them according to the definition of contextual bisimulation. 
Suppose first 
\[
\new{\vc{t}}{(P_1 \mid O)} \mid R \wact{\tau} 
\new{t_1}{(\new{t_2,\ldots,t_n}{(P_1 \mid O)} \mid 
             (\ol{c_{1}} \isum  a_{2}(t_{2})\cdots))}
\]
The reduced program cannot commit on $\ol{b}_1$ while it can commit on
$\ol{c}_1$. If $\new{\vc{t}}{(P_2 \mid O)} \mid R$ has to match this
reduction, then $R$ must necessarily perform the input action and
stop at the same point of the control  
$(\ol{c_{1}} \isum  a_{2}(t_{2})\cdots)$.
By this communication, the scope of the restricted name $t_1$ is
extruded to $R$. 
The program $O$ is composed only of emissions and therefore it cannot
change.
The program $P_2$ may perform some internal actions but it cannot
interact with $O$ and $R$.

If we repeat this argument $n$ times, we conclude that 
$\new{\vc{t}}{(P_1 \mid O)} \mid R \wact{\tau}
 \new{\vc{t}}{(P_1 \mid O \mid \ol{c_{n}} \isum s(x)\cdots)}$
and 
$\new{\vc{t}}{(P_2 \mid O)} \mid R \wact{\tau}
 \new{\vc{t}}{(P'_2 \mid O \mid \ol{c_{n}} \isum s(x)\cdots)}$
where $P_2 \wact{\tau} P'_2$.
Now the first program performs a communication on $s$ between 
$P_1$ and the residual of $R$ and, provided the emitted value
has the expected shape $\new{\vc{t'}}{v}$, it reduces to 
$\new{\vc{t},\vc{t'}}{(P'_1 \mid O \mid \ol{c_{n+1}} \isum O')}$.
In order to match this transition, it must be that 
$P'_2 \wact{\new{\vc{t'}}{\emit{s}{v}}} P''_2$ and 
the second program reduces to 
$\new{\vc{t},\vc{t'}}{(P''_2 \mid O \mid \ol{c_{n+1}} \isum O')}$.
Now if the first program moves to 
$\new{\vc{t},\vc{t'}}{(P'_1 \mid O \mid O')}$, the
second must move to 
$\new{\vc{t},\vc{t'}}{(P'''_2 \mid O \mid O')}$ where
$P''\wact{\tau} P'''_2$ and 
$\new{\vc{t},\vc{t'}}{(P'_1 \mid O \mid O')} \cbis 
 \new{\vc{t},\vc{t'}}{(P'''_2 \mid O \mid O')}$.
Since $P_2 \wact{\tau}\cdot \wact{\new{\vc{t'}}{\emit{s}{v}}} \cdot \wact{\tau} P'''_2$,
we can conclude that 
$P_2 \wact{\new{\vc{t'}}{\emit{s}{v}}} P'''_2$ and $P'_1\ \cl{R}\ P'''_2$.

\Proofitem{(L3)}
Suppose $P_1 \act{sv} P'_1$.
We consider two subcases.

\Proofitem{(L3)[1]}
Suppose $\neg P_1 \lsusp$. Then,
$\neg P'_1 \lsusp$.
By proposition \ref{char-lsusp-prop},
$\neg \new{\vc{t}}{(P_1 \mid O)}\lsusp$ and
$\neg \new{\vc{t}}{(P'_1 \mid O)}\lsusp$.
By proposition \ref{lsusp-cbis},
$\neg \new{\vc{t}}{(P_2 \mid O)}\lsusp$.
Let us show that the latter implies $\neg P_2 \lsusp$.
If $P_2\lsusp$, by proposition \ref{char-lsusp-prop}
there is a $Q$ such that $(P_2\mid Q)\wact{\tau} Q'$ and $Q' \susp$.
Then we would have:
\[
\new{\vc{t}}{(P_2\mid O)}\mid R(\vc{a})[Q] 
\wact{\tau} \new{\vc{t}}{(P_2 \mid O \mid Q)} 
\wact{\tau} \new{\vc{t}}{Q'\mid O}~.
\]
Now if $Q'\susp$ then $ \new{\vc{t}}{Q'\mid O} \susp$,
and this contradicts the hypothesis 
that $\neg \new{\vc{t}}{(P_2\mid O)} \lsusp$.
Thus $P_2 \wact{\tau} P_2$, $\neg (P_2\mid \emit{s}{v})\lsusp$,
and $P'_{1} \lbis (P_2\mid \emit{s}{v})$.

\Proofitem{(L3)[2]}
Suppose $P_1\lsusp$. In this case, the commitments are observable.
We define
\[
R=R(\vc{a})[\emit{s}{v}]
\]
Then
$\new{\vc{t}}{(P_1\mid O)}\mid R \wact{\tau}
\new{\vc{t}}{(P'_1 \mid O \mid \emit{s}{v})}$ and 
$\new{\vc{t}}{(P_2\mid O)}\mid R \wact{\tau}
\new{\vc{t}}{(P'_2 \mid O \mid \emit{s}{v})}$.
We note that 
$\new{\vc{t}}{(P'_1\mid O \mid \emit{s}{v})}\sbis 
\new{\vc{t}}{(P'_1\mid O)}$ since
$P_1\act{sv}P'_1$.
We have two subcases.

\Proofitem{(L3)[2.1]}
Suppose $P_2\wact{sv} P'_2$. Then 
$P'_2 \sbis (P'_2 \mid \emit{s}{v})$ and therefore
$P'_1 \rl{R} P'_2$ up to $\sbis$.

\Proofitem{(L3)[2.2]}
Suppose $P_2 \wact{\tau} P'_2$. Then
$P'_1 \rl{R} (P'_2\mid \emit{s}{v})$ up to
$\sbis$.

\Proofitem{(L4)}
Suppose $(P_1\mid S)\susp$ and $(P_1\mid S) \mapsto P'_1$.
We consider
\[
R_1 = R(\vc{a})[S] 
\qquad 
R_2 = R(\vc{a})[S\mid \s{pause}.O]
\]
By $(C1)$,
$\new{\vc{t}}{(P_1\mid O)}\mid R_i \cbis
\new{\vc{t}}{(P_2\mid O)}\mid R_i$ for $i=1,2$.
Also
\[
\new{\vc{t}}{(P_1\mid O)}\mid R_1 
\wact{\tau}  \new{\vc{t}}{(P_1\mid O \mid S)} \susp
\]
and 
\[
\new{\vc{t}}{(P_1\mid O)}\mid R_2 
\wact{\tau}  \new{\vc{t}}{(P_1\mid O \mid S \mid \s{pause}.O)}
\mapsto 
\new{\vc{t}}{(P'_1\mid O)}~.
\]
Then we must have:

\Defitem{(1)} $\new{\vc{t}}{(P_2\mid O)}\mid R_1 \wact{\tau}
\new{\vc{t}}{(P''_2 \mid O \mid S)}\susp$ and 
$\new{\vc{t}}{(P_1\mid O \mid S)} \cbis
\new{\vc{t}}{(P''_2\mid O \mid S)}$.
By definition of $O$ and $R_1$ this implies that
$(P_2\mid S) \wact{\tau} (P''_{2} \mid S)$ and
$(P''_{2} \mid S)\susp$.

\Defitem{(2)} $\new{\vc{t}}{(P_2\mid O)}\mid R_2 \wact{\tau}
\new{\vc{t}}{(P''_2 \mid O \mid S \mid \s{pause}.O)}
\mapsto \new{\vc{t}}{(P'_2 \mid O)}$ 
and  $\new{\vc{t}}{(P'_1\mid O)} \cbis
 \new{\vc{t}}{(P'_2 \mid O)}$. Again by definition of 
$O$ we have that $(P''_{2}\mid S) \mapsto P'_{2}$. 
\qed

\section{Conclusion}
We have proposed a {\em synchronous} version of the $\pi$-calculus
which borrows the notion of instant from the SL model--a relaxation
of the {\sc Esterel} model.
We have shown that the resulting language is amenable to a semantic
treatment similar to that available for the $\pi$-calculus.
Retrospectively, we feel that the developed theory relies on two key
insights: the introduction of the notion of L-suspension and 
the remark that the observation of signals is similar to the
observation of channels with asynchronous communication.

\end{document}